\documentclass[12pt]{article}
\usepackage{amsmath,epsfig,amssymb,bbm,cite}

\renewcommand{\Im}{{\rm Im}}

\begin{document}

\title{\vbox{
\baselineskip 14pt
\hfill \hbox{\normalsize KUNS-2256}\\
\hfill \hbox{\normalsize YITP-10-11} } \vskip 2cm
\bf Three-generation Models from $E_{8}$ Magnetized Extra Dimensional Theory
 \vskip 0.5cm
}
\author{
Tatsuo~Kobayashi$^{1}$, \ 
Ryosuke~Maruyama$^{1}$, 
Masaki~Murata$^{2}$, \   \\
Hiroshi~Ohki$^{1,2}$,  \ Manabu~Sakai$^{2}$\\*[20pt]
$^1${\it \normalsize
Department of Physics, Kyoto University,
Kyoto 606-8502, Japan} \\
$^2${\it \normalsize 
Yukawa Institute for Theoretical Physics, Kyoto University, 
Kyoto 606-8502, Japan}
}

\date{}

\maketitle
\thispagestyle{empty}

\begin{abstract}
We study 10D super Yang-Mills $E_8$ theory on the 
6D torus compactification with magnetic fluxes.
We study systematically the possibilities for 
realizing 4D supersymmetric standard models 
with three generations of quarks and leptons.
We also study quark mass matrices.
\end{abstract}

\newpage

\setcounter{page}{1}

\section{Introduction}

Extra dimensional field theories, 
in particular string-derived extra dimensional field 
theories, play an important role in particle phenomenology 
as well as cosmology.
Realization of a 4D chiral theory is one of important issues
when we start with higher dimensional theories.
Non-trivial gauge and geometrical backgrounds would 
lead to various 4D chiral theories.

Introducing magnetic fluxes is one of interesting ways 
to realize a 4D chiral theory.
Indeed, several studies on models with magnetic fluxes 
have been carried out in field theories 
and superstring theories~\cite{Manton:1981es,
Witten:1984dg,Bachas:1995ik,BDL,Blumenhagen:2000wh,
Angelantonj:2000hi,CIM,Troost:1999xn,Alfaro:2006is}.
Furthermore, magnetized D-brane models are 
T-duals of intersecting D-brane models and 
within the latter framework several interesting 
models have been constructed\cite{BDL,
Blumenhagen:2000wh,
Angelantonj:2000hi,Aldazabal:2000dg,
Blumenhagen:2000ea,Cvetic:2001tj,Honecker:2004kb}\footnote{
See for a review \cite{Blumenhagen:2005mu} and references therein.}.

In extra dimensional models with magnetic fluxes, 
the number of zero-modes is determined 
by the size of magnetic fluxes.
Their wavefunction profiles are quasi-localized 
in extra dimensions.
We can compute Yukawa couplings and higher order couplings 
in 4D effective theories 
by overlap integrals of zero-mode wavefunctions~\cite{
CIM,DiVecchia:2008tm,Antoniadis:2009bg,Abe:2009dr}.
When zero-modes are quasi-localized far away each other 
in extra dimensions, 
their couplings would be suppressed.
On the other hand, when their localizing points are 
close to each other, their couplings would be large.
Thus, extra dimensional models with magnetic fluxes 
would be quite interesting from the phenomenological 
viewpoint.
In addition to the torus compactification, 
the orbifold compactification with magnetic fluxes also 
leads to various interesting models~\cite{Abe:2008fi,Abe:2009uz}.

In most of model building with magnetic fluxes, 
one has often started with $U(N)$ gauge groups.
That is a reasonable starting point from the viewpoint of 
magnetized D-brane models.
Then, several interesting models have been constructed 
as said above.
On the other hand, gauge theories with the gauge groups, 
$E_6$, $E_7$ and $E_8$ are also interesting from 
the bottom-up phenomenological viewpoints.\footnote{
These exceptional gauge groups can be derived in 
heterotic string theory, type IIB string theory with 
non-perturbative effects and F-theory.
For example, $E_8 \times E_8$ heterotic orbifold models 
lead to realistic models~\cite{Kobayashi:2004ud,Forste:2004ie,
Buchmuller:2005jr,Kim:2006hw,Lebedev:2006kn}.
See also for another heterotic models e.g. Ref.~\cite{Blumenhagen:2005ga}.}
That is, those gauge theories are interesting 
as grand unified theories in particle physics 
and quarks and leptons are involved in 
adjoint representations of those gauge groups
in group-theoretical sense.
Thus, it would be interesting to study extra dimensional 
models with magnetic fluxes and these exceptional groups.
Indeed, such a study has been carried out in \cite{Choi:2009pv}, 
showing 4D interesting effective theories with 
semi-realistic massless spectra.
In particular, the $E_8$ models have the most variety 
because the gauge group is largest.
In this paper, we study 10D $E_8$ super Yang-Mills theory 
on the 6D torus compactification with magnetic fluxes.
We systematically classify 4D effective theories with 
semi-realistic massless spectra.

This paper is organized as follows.
In section 2, we review briefly zero-modes and their 
wavefunctions and Yukawa couplings.
In section 3, we study systematically three-generation 
models.
We also study quark mass matrices in our models.
Section 4 is devoted to conclusion and discussion.

\section{Magnetized extra dimensions}

Here we give a brief review on the torus models 
with magnetic fluxes \cite{CIM}. 
We start with 10D super Yang-Mills theory, which has the gauge 
group $G$.
We denote the vector fields and gaugino fields by 
$A_M$ ($M=0,\cdots, 9$) and $\lambda$, respectively.
Its Lagrangian is written as 
\begin{eqnarray}
{\cal L} &=& 
-\frac{1}{4g^2}{\rm Tr}\left( F^{MN}F_{MN}  \right) 
+\frac{i}{2g^2}{\rm Tr}\left(  \bar \lambda \Gamma^M D_M \lambda
\right),
\end{eqnarray}
where $\Gamma^M$ is the 
gamma matrix for ten-dimensions and 
the covariant derivative $D_M$ is given as 
\begin{eqnarray}
D_M\lambda &=& \partial_M \lambda - i [A_M, \lambda],
\end{eqnarray}
where $A_M$ is the vector field.
Furthermore, the field strength $F_{MN}$ is given by 
\begin{eqnarray}
F_{MN} &=& \partial_M A_N - \partial_N A_M -i[A_M,A_N].
\end{eqnarray}

\subsection{Zero-modes on magnetized torus }

We consider the background $R^{3,1}\times (T^2)^3$,
whose coordinates are denoted by
$x_\mu$ $(\mu=0,\cdots, 3)$ for the uncompact space $R^{3,1}$
and $y_m$ $(m=4, \cdots, 9)$ for the compact space $(T^2)^3$.
We often use complex coordinations $z_d$ $(d=1,2,3)$ 
for the $d$-th torus $T^2_d$, e.g. 
$z_1=y_4+\tau_1 y_5$.
Here, $\tau_d$ denote complex structure moduli of the $d$-th 
$T^2_d$, while the area of $T^2_d$ is denoted by ${\cal A}_d$.
The periodicity on $T^2_d$ is written as 
$z_d \sim z_d +1_d$ and $z_d \sim z_d +\tau_d$.

The gaugino fields $\lambda$ and the vector fields $A_\mu$ and $A_m$ 
are decomposed as 
\begin{eqnarray}
\lambda(x,z) &=& \sum_n \chi_n(x) \otimes \psi_n(z), 
\nonumber \\
A_\mu(x,z) &=& \sum_n A_{n,\mu}(x) \otimes \phi_{n,\mu}(z),
 \\
A_m(x,z) &=& \sum_n \varphi_{n,m}(x) \otimes \phi_{n,m}(z).\nonumber
\end{eqnarray}
Hereafter, we concentrate on zero-modes, $\psi_0(z)$, and 
we denote them as $\psi(z)$ by omitting the subscript ``0''.
Furthermore, the internal part $\psi(z)$ is decomposed 
as a product of the $T^2_d$ parts, i.e. 
$\psi_{(d)}(z_d)$.
Each of $\psi_{(d)}(z_d)$ is two-component spinor, 
\begin{eqnarray}
\psi_{(d)}= \left(
\begin{array}{c}
\psi_{+(d)} \\
\psi_{-(d)}
\end{array}
\right),
\end{eqnarray}
and their chirality for the $d$-th part is denoted by $s_d$.
We use the following gamma matrix for $T^2_d$ 
\begin{eqnarray}
\tilde \Gamma^1_{(d)} = \left(
\begin{array}{cc}
0 & 1 \\ 1 & 0 
\end{array}
\right), \qquad \tilde \Gamma^2_{(d)}=\left(
\begin{array}{cc}
0 & -i \\ i & 0 
\end{array}
\right).
\end{eqnarray}

We introduce the magnetic flux along the $U(1)_a$ (Cartan) direction 
of $G$ on $T^2_d$,
\begin{eqnarray}\label{eq:mag-flux}
F  ={\pi i \over \Im \tau_d} m_{(d)}^a \ (dz_d \wedge d \bar z_d), 
\end{eqnarray}
where $m_{(d)}^a$ is an integer~\cite{toron}.
Here, we normalize $U(1)_a$ charges $q^a$ such that 
all  $U(1)_a$ charges are integers and the minimum satisfies 
$|q^a|=1$.
The above magnetic flux can be obtained from the vector potential,
\begin{eqnarray}
A(z_d) ={\pi m^a_{(d)} \over \Im \tau_d} \Im (\bar z_d \ dz_d).
\end{eqnarray}
This form of the vector potential satisfies the 
following relations,
\begin{eqnarray}
A(z_d+1_d) &=& A(z_d) +{\pi m^a_{(d)} \over \Im \tau_d} \Im (dz_d), 
\nonumber \\
A(z_d+\tau_d ) &=& A(z_d) +{\pi m^a_{(d)} \over \Im \tau_d} 
\Im (\bar \tau_d \ dz_d).
\end{eqnarray}
These relations can be represented as the following 
gauge transformations,
\begin{eqnarray}
A(z_d+1_d) = A(z_d) + d \chi_1^{(d)}, \qquad 
A(z_d+\tau_d ) = A(z_d) + d \chi_2^{(d)},
\end{eqnarray}
where 
\begin{eqnarray}\label{eq:chi}
\chi_1^{(d)} = {\pi m^a_{(d)} \over \Im \tau_d} \Im (z_d), \qquad 
\chi_2^{(d)} = {\pi m^a_{(d)} \over \Im \tau_d} \Im (\bar \tau_d \ z_d).
\end{eqnarray}
Then, the fermion field $\psi_{(d)} (z_d)$ with the $U(1)_a$ charge $q^a$ must 
satisfy 
\begin{eqnarray}
\psi_{(d)}  (z_d+1_d) = e^{iq^a\chi_1^{(d)}(z_d)} \psi_{(d)} (z_d), \qquad 
\psi_{(d)}  (z_d+\tau_d) = e^{iq^a\chi_2^{(d)}(z_d)} \psi_{(d)} (z_d).
\end{eqnarray}

By the magnetic flux (\ref{eq:mag-flux}) along the $U(1)_a$ direction,  
all of 4D gauge vector fields $A_\mu$ with non-vanishing $U(1)_a$ charges, 
become massive, that is, the gauge group is broken from 
$G$ to $G'\times U(1)_a$ without reducing its rank,\footnote{For example, when 
$G=SU(N)$, $G'$ would correspond to $SU(N-1)$.} where 
4D gauge fields $A_\mu$ in $G' \times U(1)_a$ have vanishing 
$U(1)_a$ charges and their zero-modes $\phi_\mu(z)$ have 
a flat profile.
Since the magnetic flux has no effect on the unbroken 
gauge sector, 4D N=4 supersymmetry remains in the 
$G'\times U(1)_a$ sector, that is, there are massless 
four adjoint gaugino fields and six adjoint scalar fields.\footnote{
In string terminology, these adjoint scalar fields correspond to 
open string moduli, that is, D-brane position moduli.
How to stabilize these moduli is one of important issues.}

In addition, matter fields appear from  gaugino fields 
corresponding to the broken gauge part, that is, 
they have non-trivial representations under $G'$ and non-vanishing
$U(1)_a$ charges $q^a$.
The Dirac equations for their zero-modes become 
\begin{eqnarray}
& & \left( \bar \partial_{z_d} + 
\frac{\pi  q^am^a_{(d)}}{2\Im (\tau_d)}  z_d \right) 
\psi_{+(d)}(z_d,\bar z_d) =0, \\
& & \left( \partial_{z_d} - 
\frac{\pi  q^am^a_{(d)}}{2\Im (\tau_d)} \bar z_d \right) 
\psi_{-(d)} (z_d,\bar z_d) =0 ,
\end{eqnarray}
for $T^2_d$.
When $q^am^a_{(d)} >0$, the component $\psi_{+(d)}$ has $M=q^am^a_{(d)}$ 
independent zero-modes and their wavefunctions are  written as \cite{CIM}
\begin{eqnarray}\label{eq:wf}
\Theta^{j,M}(z)=N_M e^{i \pi Mz\Im (z)/ \Im (\tau)} \vartheta 
\left[
\begin{array}{c}
j/M \\ 0
\end{array}
\right]
\left( Mz, M \tau \right),
\end{eqnarray}
where 
$j$ denotes the flavor index, i.e. 
$j=1,\cdots, M$ and 
\begin{eqnarray}
\vartheta \left[ 
\begin{array}{c}
a \\ b
\end{array} \right]
\left( \nu, \mu \right) 
&=& 
\sum_n \exp\left[ \pi i
   (n+a)^2 \mu + 2 \pi i (n+a)(\nu +b)\right],
\nonumber
\end{eqnarray}
that is, the Jacobi theta-function.
Here, the normalization factor $N_M$ is obtained 
as
\begin{equation} \label{normalization}
N_{M^{(d)}} = \left( {2 \Im \tau_{d} |M^{(d)}| \over
   A_{d}^2 } \right)^{1/4} .
\end{equation}
Note that $\Theta^{0,M}(z)=\Theta^{M,M}(z)$.
Furthermore, for $q^am^a_{(d)} > 0$, the other component 
$\psi_{-(d)}$ has no zero-modes.
On the other hand, when $q^am^a_{(d)} <0$, 
the component $\psi_{-(d)}$ has 
$|q^am^a_{(d)} |$ independent zero-modes, but the other component 
$\psi_{+(d)}$ has no zero-modes.

As a result, we can realize a chiral spectrum when 
we introduce magnetic fluxes on all of three $T^2_d$.
That is, since the ten-dimensional chirality of gaugino fields is fixed, 
zero-modes for either $q^a > 0$ and  $q^a < 0$ appear 
with a fixed four-dimensional chirality.

We can also introduce Wilson lines along the $U(1)_b$ direction of 
$G'$.
That breaks further the gauge group $G'$ to $G'' \times
U(1)_b$ without reducing its rank.\footnote{
For example, when $G'=SU(N-1)$, the Wilson line breaks it 
to $SU(N-2)\times U(1)_b$.}
All of the $U(1)_b$-charged fields including 4D vector, spinor and 
scalar fields become massive because of the Wilson line, when 
they are not charged under $U(1)_a$ and their zero-mode profiles 
are flat.
On the other hand, the matter fields with non-trivial profiles 
due to magnetic flux have different behavior. 
For matter fields with $U(1)_a$ charge $q^a$ and $U(1)_b$ charge $q^b$, 
the Dirac equations of the zero-modes are modified 
by the Wilson line background, $C^b_d = C^b_{d,1} + \tau_d C^b_{d,2}$ as 
\begin{eqnarray}\label{eq:zero-mode-WL-b}
& & \left( \bar \partial_{z_d} + \frac{\pi  }{2\Im (\tau_d)} 
(q^am^a_{(d)}z_d+q^bC^b_d) \right) 
\psi_{+(d)}(z_d,\bar z_d) =0, \\
& & \left( \partial_{z_d} - \frac{\pi  }{2\Im (\tau_d)} 
 (q^am^a_{(d)}\bar z_d +q^b\bar C^b_d) \right) 
\psi_{-(d)} (z_d,\bar z_d) =0,
\end{eqnarray}
where $C^b_{d,1}$ and $C^b_{d,2}$ are real parameters.
That is, we can introduce Wilson lines along the $U(1)_b$ 
direction by replacing $\chi_i^{(d)}$ in 
(\ref{eq:chi}) as~\cite{CIM}
\begin{eqnarray}\label{eq:chi-WL}
\chi_1^{(d)} = {\pi  \over \Im \tau_d} 
\Im (m^a_{(d)}z_d +q^bC^b_d/q^a), \qquad 
\chi_2^{(d)} = {\pi  \over \Im \tau_d} 
\Im (\bar \tau_d (m^a_{(d)}z_d + q^bC^b_d/q^a)).
\end{eqnarray}
Because of this Wilson line, the number of 
zero-modes does not change, but their wave functions 
are shifted as 
\begin{eqnarray}\label{eq:WL-shift}
\Theta^{j,M}(z_d) \rightarrow \Theta^{j,M}(z_d+q^bC^b_d/(q^am^a_{(d)})).
\end{eqnarray}
Note that the shift of zero-mode profiles depend on 
$U(1)_b$ charges of matter fields.
We often denote the degree of Wilson lines as 
$\zeta^{(d)}= q^bC^b_d/(q^am^a_{(d)})$.
Also, we can introduce the Wilson line $C^a_d$ 
along the $U(1)_a$ direction.

Similarly we can analyze 4D massless scalar modes~\cite{CIM}.
When the above magnetic fluxes satisfies the following relation,
\begin{eqnarray}\label{eq:SUSY-condition}
\sum_{d=1}^3 \pm  \frac{m^a_{(d)}}{{\cal A}_d} = 0,
\end{eqnarray}
for one combinations of signs, 
the 4D N=1 supersymmetry is 
preserved~\cite{Troost:1999xn,CIM,Antoniadis:2004pp} 
for the above fermion fields with the $U(1)$ charge $q_a$.
That is, there is the same number of 4D scalar zero-modes 
and their wave function profiles are the same as 
the above fermion fields.
For example, for Higgs fields, we study 
zero-modes and their profiles of Higgsino fields.

\subsection{Yukawa couplings}

The Yukawa couplings of zero-modes 
in 4D effective theory can be computed by 
overlap integral of their wavefunctions.
For example, we consider the coupling among three fields, 
whose wavefunctions are written by 
$\psi^{i}(z)$, $\psi^{j}(z)$ and $\left( \psi^{k}(z)
\right)^*$.
Their 4D Yukawa couplings are obtained by 
the overlap integral  of wavefunctions
\begin{equation}
  Y_{ij\bar k} = g\int d^6 z \ 
\psi^{i}(z) \psi^{j}(z) \left( \psi^{k}(z) \right)^* ,
\end{equation}
on the 6D extra dimensions.

Suppose that the matter fields $\psi^{i}(z)$, $\psi^{j}(z)$ 
and $\left( \psi^{k}(z) \right)^*$ have 
$M_1^{(d)}$, $M_2^{(d)}$ and  $M_3^{(d)}$ zero-modes 
on the $d$-th torus $T^2_{(d)}$ by certain magnetic fluxes and 
their wavefunctions are affected by Wilson lines, 
$\zeta^{(d)}_1$, $\zeta^{(d)}_2$ and $\zeta^{(d)}_3$ 
on the  $d$-th torus $T^2_{(d)}$.
Then, their wavefunctions on $T^2_{(d)}$ are 
written by $\Theta^{i,M^{(d)}_1}(z_d+\zeta^{(d)}_1)$, 
$\Theta^{j,M^{(d)}_2}(z_d+\zeta^{(d)}_2)$ 
and $\left(\Theta^{k,M^{(d)}_3}(z_d+\zeta^{(d)}_3)\right)^*$
as Eq.~(\ref{eq:wf}).
In this case, the corresponding Yukawa coupling can 
be written as \cite{CIM}
\begin{equation} \label{Yukawa} \begin{split}
 Y_{ij \bar k} = & g \prod_{d=1}^3 \left({2 \Im \tau_d \over
   A^2_d}{M_1^{(d)} M_2^{(d)} \over M_3^{(d)}}\right)^{1/4} \\
 &\times \exp\left(i \pi (M_1^{(d)} \zeta_1^{(d)} \Im \zeta_1^{(d)} + M_2^{(d)} \zeta_2^{(d)} \Im \zeta_2^{(d)} + M_3^{(d)}
   \zeta_3^{(d)} \Im \zeta_3^{(d)})/\Im \tau_d \right)
\\   &\times  \vartheta \left[ \begin{matrix}
    \frac{i}{ M_1^{(d)}} + \frac{j}{M_2^{(d)}} +  \frac{k}{M_3^{(d)}} 
   \\ 0 \end{matrix} 
  \right](\tilde \zeta^{(d)},\tau_d M_1^{(d)} M_2^{(d)} M_3^{(d)}) ,
 \end{split} \end{equation}
where 
$\tilde \zeta^{(d)} = M_2^{(d)} M_3^{(d)}(\zeta_2^{(d)} - \zeta_3^{(d)})$.

\section{$E_8$ theory}

Here, we study 10D N=1 super Yang-Mills theory with 
the gauge group $E_8$.

\subsection{Magnetic fluxes}

When we decompose $E_8$ to $SU(3) \times SU(2) \times U(1)_Y 
\times U(1)^4$, $U(1)^5$ including $U(1)_Y$ appear.  
We can introduce magnetic fluxes along these $U(1)^5$ directions.
The ${\bf 248}$ adjoint representation of $E_8$ is 
decomposed to several representations under $SU(3) \times SU(2) \times U(1)_Y 
\times U(1)^4$.
Certain representations are shown in Table \ref{tab:248}, 
where we follow the notation in Ref.~\cite{Bourjaily:2009ci}.
The total ${\bf 248}$ representation consists of 
the representations in Table \ref{tab:248}, their 
conjugate representations and the adjoint representations 
of $SU(3) \times SU(2) \times U(1)_Y \times U(1)^4$.

Now, we introduce magnetic fluxes $m^I_{(d)}$ 
along five  $U(1)_I$ ($I=a,b,c,d,Y$) directions 
on the $d$-torus.
Then, the sum of magnetic fluxes $\sum_I q^Im^I_{(d)}$ 
appears in the zero-mode Dirac equation for the matter fields 
with the $U(1)_I$ charges $q^I$.
We require $\sum_I q^Im^I_{(d)}$  to be integers for all of matter 
fields, that is, the quantization conditions of 
magnetic fluxes.
For example, five $({\bf 3},{\bf 2})_{1}$ representations 
under $SU(3) \times SU(2) \times U(1)_Y$ 
as well as their conjugates appear from the ${\bf 248}$ adjoint 
representation.
In the zero-mode equations of these five $({\bf 3},{\bf 2})_{1}$ 
matter fields, $Q_i$ ($i=1,\cdots,5$), the following sum of 
magnetic fluxes $\sum_I q^Im^I_{(d)}$ appear 
\begin{eqnarray}
m^{Q1}_{(d)} &=& m^a_{(d)} + m^c_{(d)} - m^d_{(d)} + m^{Y}_{(d)}, 
\nonumber \\
m^{Q2}_{(d)} &=& m^b_{(d)} + m^c_{(d)} - m^d_{(d)} + m^{Y}_{(d)}, 
\nonumber \\
m^{Q3}_{(d)} &=& -m^a_{(d)} -m^b_{(d)} 
+ m^c_{(d)} - m^d_{(d)} + m^{Y}_{(d)}, \\
m^{Q4}_{(d)} &=& -3 m^c_{(d)} - m^d_{(d)} + m^{Y}_{(d)}  \nonumber \\
m^{Q5}_{(d)} &=& 4m^d_{(d)} + m^{Y}_{(d)} .   \nonumber
\end{eqnarray}
Thus, we have to take the magnetic fluxes such that 
all of $m^{Qi}_{(d)}$ for $i=1,\cdots, 5$ are integers.

We can write the sum of magnetic fluxes $\sum_I q^Im^I_{(d)}$ 
for the other matter fields in terms of $m^{Qi}_{(d)}$.
For example, the sum of magnetic fluxes $\sum_I q^Im^I_{(d)}$ 
for the matter field $Q^c_Y$ can be written as 
\begin{eqnarray}
m^{QY}_{(d)} &=& m^{Q1}_{(d)} + m^{Q2}_{(d)} +m^{Q3}_{(d)} +
m^{Q4}_{(d)} + m^{Q5}_{(d)}.
\end{eqnarray}
Thus, if all of $m^{Qi}_{(d)}$ for $i=1,\cdots, 5$ are integers, 
$m^{QY}_{(d)}$ are also integers.
Similarly, the sums of magnetic fluxes $\sum_I q^Im^I_{(d)}$  
are written as 
\begin{eqnarray}
m^{ui}_{(d)} &=& m^{Qi}_{(d)}- m^{QY}_{(d)},~~~~~~~~~~~~{\rm for}~~u^c_i \ (i=1,\cdots,5),
\nonumber \\
m^{ei}_{(d)} &=& m^{Qi}_{(d)}+ m^{QY}_{(d)},~~~~~~~~~~~~{\rm for}~~e^c_i \ (i=1,\cdots,5),
\nonumber \\
m^{di}_{(d)} &=& m^{Qi}_{(d)}+ m^{Q5}_{(d)},~~~~~~~~~~~~~{\rm for}~~d^c_i \ (i=1,\cdots,4),
\nonumber \\
m^{Li}_{(d)} &=& m^{Qi}_{(d)}+ m^{Q5}_{(d)} -m^{QY}_{(d)},~~~{\rm
  for}~~L_i \ (i=1,\cdots,4),
\\
m^{\nu i}_{(d)} &=& m^{Qi}_{(d)}- m^{Q5}_{(d)},~~~~~~~~~~~~~{\rm for}~~\nu^c_i \ 
(i=1,\cdots,4),
\nonumber \\
m^{S i}_{(d)} &=& m^{Qi}_{(d)}- m^{Q4}_{(d)},~~~~~~~~~~~~~{\rm for}~~S_i \ 
(i=1,2,3) , \nonumber  \\
m^{D^c i}_{(d)} &=& m^{Qi}_{(d)}+ m^{Q4}_{(d)},~~~~~~~~~~~~~{\rm for}~~D^c_i \ 
(i=1,2,3) . \nonumber 
\end{eqnarray}
In addition, the sums $\sum_I q^Im^I_{(d)}$  
are written as 
\begin{eqnarray}
m^{D1}_{(d)}=-m^{Q2}_{(d)}-m^{Q3}_{(d)}, \qquad 
m^{D2}_{(d)}=-m^{Q3}_{(d)}-m^{Q1}_{(d)}, \qquad 
m^{D3}_{(d)}=-m^{Q1}_{(d)}-m^{Q2}_{(d)}, 
\end{eqnarray}
for the matter fields $D_1$, $D_2$ and $D_3$, 
\begin{eqnarray}
m^{N1}_{(d)}=-m^{Q1}_{(d)}-m^{Q3}_{(d)}, \qquad 
m^{N2}_{(d)}=-m^{Q2}_{(d)}-m^{Q3}_{(d)}, \qquad 
m^{N3}_{(d)}=-m^{Q1}_{(d)}-m^{Q2}_{(d)}, 
\end{eqnarray}
for the matter fields $N_1$, $N_2$ and $N_3$.
Also the sums $\sum_I q^Im^I_{(d)}$  
are written as 
\begin{eqnarray}
m^{H^u i}_{(d)} &=& m^{Di}_{(d)}+ m^{QY}_{(d)},~~~{\rm for}~~H^u_i \ 
(i=1,2,3) , \nonumber  \\
m^{H^d i}_{(d)} &=& m^{D^c i}_{(d)}- m^{QY}_{(d)},~~~{\rm for}~~H^d_i \ 
(i=1,2,3) . 
\end{eqnarray}
Note that the sums $\sum_I q^Im^I_{(d)}$ for all of 
matter fields can be written in terms of $m^{Qi}_{(d)}$
($i=1,\cdots,5$) with integer coefficients.
Hence, when all of $m^{Qi}_{(d)}$ are integers, 
the sums $\sum_I q^Im^I_{(d)}$ for the other matter fields 
are always integers.
Hereafter, we show the magnetic fluxes in terms of $m^{Qi}_{(d)}$
($i=1,\cdots,5$).
That is, we classify models by studying systematically 
combinations of $m^{Qi}_{(d)}$ ($i=1,\cdots,5$).
We use the notation $m^\Phi=\prod_{d=1}^3m^\Phi_{(d)}$ for 
the matter field $\Phi$ 
and it denotes the total zero-mode numbers of $\Phi$.
When $m^\Phi < 0$, the matter fields with conjugate representations 
appear, i.e. the anti-generations of $\Phi$.

\subsection{Three-generation models}

If the condition (\ref{eq:SUSY-condition}) is 
satisfied, 4D N=1 supersymmetry is preserved and 
tachyonic modes do not appear.
Thus, first we concentrate on combinations of $m^{Qi}_{(d)}$
($i=1,\cdots,5$), which satisfy the condition
(\ref{eq:SUSY-condition}).
Here, we consider the area ${\cal A}_d$ as free parameters.
Only their ratios, e.g. 
${\cal A}_2/{\cal A}_1$ and ${\cal A}_3/{\cal A}_1$, are 
important to satisfy the condition (\ref{eq:SUSY-condition}).
That is, there are two free parameters.
Thus, all of five vectors  
$(m^{Qi}_{(1)},m^{Qi}_{(2)},m^{Qi}_{(3)})$ $(i=1,\cdots, 5)$, 
are not independent of each other  
to satisfy the condition (\ref{eq:SUSY-condition}).
However, when five vectors  
$(m^{Qi}_{(1)},m^{Qi}_{(2)},m^{Qi}_{(3)})$ $(i=1,2,3)$ are written 
in terms of two (independent) vectors, 
we can choose areas ${\cal A}_d$ such that they satisfy 
the condition (\ref{eq:SUSY-condition}).
Furthermore, a tachyonic mode would appear 
if one of $m^{\Phi}_{(i)}$ $(i=1,2,2)$ is not 
vanishing and the other two are vanishing, e.g. 
$(m^{\Phi}_{(1)},m^{\Phi}_{(2)},m^{\Phi}_{(3)}) =(m',0,0)$ with 
$m'\neq 0$. 
We rule out such a case.

In addition, we concentrate our systematic study on 
the following regions of $m^{Qi}_{(d)}$
($i=1,\cdots,5$).
The matter fields $Q_i$ would correspond to left-handed 
quark doublets.
Thus, we require that there are three generations, i.e.
\begin{eqnarray}\label{eq:condition-Q1}
\sum_{i=1}^{5} m^{Qi} =3.
\end{eqnarray}
On top of that, we concentrate on 
\begin{eqnarray}\label{eq:condition-Q2}
m^{Qi}  \geq 0
\end{eqnarray}
for each $Q_i$.
That means that there is no anti-generations 
for quark doublets.

For the other matter fields, we allow anti-generations, but 
we require the total numbers of (chiral) generations to be equal to 
three, 
\begin{eqnarray}\label{eq:condition-u-3}
& & \sum_{i=1}^{5} m^{ui} =3, \qquad \sum_{i=1}^{5} m^{ei} =3, 
\qquad \sum_{i=1}^{4} m^{di} =3,  \\
& & \sum_{i=1}^4 m^{Li} + \sum_{i=1}^3 m^{H^di} -   \sum_{i=1}^3
m^{H^ui} = 3.
\label{eq:condition-L}
\end{eqnarray}
Only by the representations under $SU(3) \times SU(2) \times U(1)_Y$, 
one can not distinguish $L_i$, $H^d_i$ and conjugates of $H^u_i$.
Thus, the last equation means that the number of 
the total chiral generation for the matter fields $({\bf 1},{\bf
  2})_{-3}$ under $SU(3) \times SU(2) \times U(1)_Y$ is equal to three
and there are some vector-like generations with such a representation,
which would correspond to pairs of Higgino fields.
For simplicity, we concentrate $m^{Di}=m^{D^ci}=0$.
The matter field $Q_Y$ has a representation similar to 
$Q_i$, but its $U(1)_Y$ charge is different.
Thus, this field would correspond to an exotic matter field,
and we require 
\begin{eqnarray}\label{eq:condition-QY}
m^{QY}=0.
\end{eqnarray}
We do not put any constraints on  the  
$SU(3) \times SU(2) \times U(1)_Y$ singlets.
We will classify the three-generation models with 
the above conditions in what follows.

All possible combinations are classified into 
the following seven types, 
\begin{eqnarray}\label{eq:7-types}
{\rm I}: & &  (m^{Q1},m^{Q2},m^{Q3},m^{Q4},m^{Q5}) = (1,1,1,0,0), \nonumber \\
{\rm II}: & &  (m^{Q1},m^{Q2},m^{Q3},m^{Q4},m^{Q5}) = (1,1,0,0,1), \nonumber \\
{\rm III}: & &  (m^{Q1},m^{Q2},m^{Q3},m^{Q4},m^{Q5}) = (2,0,0,0,1), \nonumber \\
{\rm IV}: & &  (m^{Q1},m^{Q2},m^{Q3},m^{Q4},m^{Q5}) = (1,0,0,0,2), \\
{\rm V}: & &  (m^{Q1},m^{Q2},m^{Q3},m^{Q4},m^{Q5}) = (2,1,0,0,0), \nonumber \\
{\rm VI}: & &  (m^{Q1},m^{Q2},m^{Q3},m^{Q4},m^{Q5}) = (0,0,0,0,3), \nonumber \\
{\rm VII}: & &  (m^{Q1},m^{Q2},m^{Q3},m^{Q4},m^{Q5}) = (3,0,0,0,0). \nonumber 
\end{eqnarray}
Note that the matter representations in Table \ref{tab:248} 
have the permutation symmetries among $(Q_i,u^c_i,e^c_i,d^c_i,L_i)$ 
for $i=1,2,3$.
Furthermore, our conditions for the three-generation models 
are symmetric under the permutations among $m^{Qi}_{(d)}$ for 
$i=1,2,3,4$.
Up to such permutation symmetries, each of possible combinations 
is equivalent to one of the above types.

First, let us study the type I in (\ref{eq:7-types}).
We choose 
\begin{eqnarray}\label{eq:type-1-1}
(m^{Q1}_{(1)},m^{Q1}_{(2)},m^{Q1}_{(3)})=(1,1,1).
\end{eqnarray}
Then, we can not take $m^{Q2}_{(d)}=m^{Q1}_{(d)}$, 
because that leads to $m^{D3} \neq 0$.
Thus, the possible values of $m^{Q2}_{(d)}$ are 
\begin{eqnarray}\label{eq:type-1-2}
(m^{Q2}_{(1)},m^{Q2}_{(2)},m^{Q2}_{(3)})=(1,-1,-1),
\end{eqnarray}
and permutations of the entries.
Similarly, for (\ref{eq:type-1-1}) and (\ref{eq:type-1-2}), 
the condition leads to 
\begin{eqnarray}
(m^{Q3}_{(1)},m^{Q3}_{(2)},m^{Q3}_{(3)})=(-1,-1,1),\quad 
(-1,1,-1).
\end{eqnarray}
However, these vectors, 
$(m^{Qi}_{(1)},m^{Qi}_{(2)},m^{Qi}_{(3)})$ $(i=1,\cdots, 3)$, 
are independent of each other.
Then, we can not find ${\cal A}_d$, which satisfy the 
SUSY condition.
Thus, the type I is not interesting.

Similarly, we can study the type II in (\ref{eq:7-types}).
We chose the same $m^{Q1}_{(d)}$ and $m^{Q2}_{(d)}$
as (\ref{eq:type-1-1}) and (\ref{eq:type-1-2}).
Both $m^{Q3}_{(d)}$ and $m^{Q4}_{(d)}$ must be written by 
linear combinations of $m^{Q1}_{(d)}$ and $m^{Q2}_{(d)}$.
Also the products, $\prod_d m^{Q3}_{(d)}$ and  $\prod_d m^{Q4}_{(d)}$, 
must vanish.
Then, possible combinations are obtained as 
\begin{eqnarray}
 & & (m^{Q3}_{(1)},m^{Q3}_{(2)},m^{Q3}_{(3)})=(0,2m,2m),\quad 
(m^{Q4}_{(1)},m^{Q4}_{(2)},m^{Q4}_{(3)})=(0,2n,2n),  \nonumber \\
 & & (m^{Q3}_{(1)},m^{Q3}_{(2)},m^{Q3}_{(3)})=(0,2m,2m),\quad 
(m^{Q4}_{(1)},m^{Q4}_{(2)},m^{Q4}_{(3)})=(2n,0,0),   \nonumber \\
 & & (m^{Q3}_{(1)},m^{Q3}_{(2)},m^{Q3}_{(3)})=(2m,0,0),\quad 
(m^{Q4}_{(1)},m^{Q4}_{(2)},m^{Q4}_{(3)})=(0,2n,2n),   \\
 & & (m^{Q3}_{(1)},m^{Q3}_{(2)},m^{Q3}_{(3)})=(2m,0,0),\quad 
(m^{Q4}_{(1)},m^{Q4}_{(2)},m^{Q4}_{(3)})=(2n,0,0),  \nonumber 
\end{eqnarray}
where $m$ and $n$ are integers.
The first three combinations do not satisfy 
$m^{D1}=-\prod (m^{Q2}_{(d)}+m^{Q3}_{(d)}) =0$ and 
$m^{D2}=-\prod (m^{Q1}_{(d)}+m^{Q3}_{(d)}) =0$.
The last combination satisfies $m^{D1}=m^{D2}=0$ 
when $m=n=-1$.
However, such a case does not lead to
$m^{QY}=\prod_d \left(\sum_{i=1}^5m^{Qi}_{(d)}\right) =0$, 
because $\sum_{i=1}^4 m^{Qi}_{(d)} =0$ and 
$m^{Q5}_{(d)} \neq 0$ for any $d$.
In addition, the last three combinations lead to tachyonic modes.
Thus, the type II does not lead to three-generation models.

We study the type V in  (\ref{eq:7-types}).
We choose 
\begin{eqnarray}\label{eq:type-5-1}
(m^{Q1}_{(1)},m^{Q1}_{(2)},m^{Q1}_{(3)})=(2,1,1).
\end{eqnarray}
For $m^{Q2}_{(d)}$, we have two possibilities,
\begin{eqnarray}\label{eq:type-5-1}
(m^{Q2}_{(d)},m^{Q2}_{(2)},m^{Q2}_{(3)})=(1,-1,-1), \quad 
(-1,-1,1).
\end{eqnarray}
Then, we require other magnetic fluxes 
$(m^{Qi}_{(1)},m^{Qi}_{(2)},m^{Qi}_{(3)})$ for $i=3,4,5$
can be written by linear combinations of 
$(m^{Q1}_{(1)},m^{Q1}_{(2)},m^{Q1}_{(3)})$ 
and $(m^{Q2}_{(1)},m^{Q2}_{(2)},m^{Q2}_{(3)})$ with 
integer coefficients.
However, any combinations of this type can not lead to 
the three-generation models.
For example, the conditions 
$m^{D1}=-\prod (m^{Q2}_{(d)}+m^{Q3}_{(d)}) =0$ and 
$m^{D2}=-\prod (m^{Q1}_{(d)}+m^{Q3}_{(d)}) =0$ are not 
satisfied.

The situation in the type III of (\ref{eq:7-types}) 
is similar.
We take the same $(m^{Q1}_{(1)},m^{Q1}_{(2)},m^{Q1}_{(3)})$ 
as Eq.~(\ref{eq:type-5-1}).
We have three possibilities for $m^{Q5}_{(d)}$ as 
\begin{eqnarray}\label{eq:type-5-2}
(m^{Q5}_{(d)},m^{Q5}_{(2)},m^{Q5}_{(3)})=(1,1,1), \quad (1,-1,-1), \quad 
(-1,-1,1).
\end{eqnarray}
Then, we require other magnetic fluxes 
$(m^{Qi}_{(1)},m^{Qi}_{(2)},m^{Qi}_{(3)})$ for $i=2,3,4$
can be written by linear combinations of 
$(m^{Q1}_{(1)},m^{Q1}_{(2)},m^{Q1}_{(3)})$ 
and $(m^{Q5}_{(1)},m^{Q5}_{(2)},m^{Q5}_{(3)})$ with 
integer coefficients.
However, any combinations of this type can not lead to 
three-generation models.

The situation in the type IV of (\ref{eq:7-types}) 
is similar.
We choose 
\begin{eqnarray}\label{eq:type-4-1}
(m^{Q1}_{(1)},m^{Q1}_{(2)},m^{Q1}_{(3)})=(1,1,1),
\end{eqnarray}
and we have three possibilities for $m^{Q5}_{(d)}$ as 
\begin{eqnarray}\label{eq:type-4-2}
(m^{Q5}_{(d)},m^{Q5}_{(2)},m^{Q5}_{(3)})=(2,1,1), \quad (2,-1,-1), \quad 
(-2,-1,1).
\end{eqnarray}
Then, we require other magnetic fluxes 
$(m^{Qi}_{(1)},m^{Qi}_{(2)},m^{Qi}_{(3)})$ for $i=2,3,4$
can be written by linear combinations of 
$(m^{Q1}_{(1)},m^{Q1}_{(2)},m^{Q1}_{(3)})$ 
and $(m^{Q5}_{(1)},m^{Q5}_{(2)},m^{Q5}_{(3)})$ with 
integer coefficients.
However, any combinations of this type can not lead to 
three-generation models.

Now, let us study the type VI in (\ref{eq:7-types}).
We choose 
\begin{eqnarray}\label{eq:type-6-1}
(m^{Q5}_{(1)},m^{Q5}_{(2)},m^{Q5}_{(3)})=(3,1,1).
\end{eqnarray}
Other magnetic fluxes $m^{Qi}_{(d)}$ for $i=1,2,3,4$ must 
have vanishing elements for one of $d=1,2,3$.
Various combinations are possible as 
\begin{eqnarray}
 & & (m^{Q1}_{(1)},m^{Q2}_{(1)},m^{Q3}_{(1)},m^{Q4}_{(1)}) =
 (0,0,0,0), \nonumber  \\
 & & (m^{Q1}_{(1)},m^{Q2}_{(1)},m^{Q3}_{(1)},m^{Q4}_{(2)}) =
 (0,0,0,0),  \nonumber   \\
 & & (m^{Q1}_{(1)},m^{Q2}_{(1)},m^{Q3}_{(2)},m^{Q4}_{(2)}) =
 (0,0,0,0), \nonumber  \\
 & & (m^{Q1}_{(1)},m^{Q2}_{(1)},m^{Q3}_{(2)},m^{Q4}_{(3)}) =
 (0,0,0,0), \nonumber  \\
 & & (m^{Q1}_{(1)},m^{Q2}_{(2)},m^{Q3}_{(2)},m^{Q4}_{(2)}) = (0,0,0,0), \\
 & & (m^{Q1}_{(1)},m^{Q2}_{(2)},m^{Q3}_{(2)},m^{Q4}_{(3)}) = 
(0,0,0,0), \nonumber  \\
 & & (m^{Q1}_{(2)},m^{Q2}_{(2)},m^{Q3}_{(2)},m^{Q4}_{(2)}) = 
(0,0,0,0), \nonumber  \\
 & & (m^{Q1}_{(2)},m^{Q2}_{(2)},m^{Q3}_{(2)},m^{Q4}_{(3)}) = 
(0,0,0,0), \nonumber  \\
 & & (m^{Q1}_{(2)},m^{Q2}_{(2)},m^{Q3}_{(3)},m^{Q4}_{(3)}) = 
(0,0,0,0). \nonumber  
\end{eqnarray}
Most of them do not lead to the three-generation models with 
the required conditions, but 
certain combinations of $m^{Qi}_{(d)}$ lead to three-generation
models.
Such combinations are shown in Tables 4, 6, 8, 10, 12, 14, 16, 18, 20 
and 22 as the model VI-1,$\cdots$,10.
In those tables, the second, third and fourth rows show 
each of magnetic fluxes on $T^2_{(1)}, T^2_{(2)}$ and $T^2_{(3)}$, 
respectively. 
The corresponding massless spectra are shown in Tables 
5, 7, 9, 11, 13, 15, 17, 19, 21 and 23.
In those table, the second column shows the zero-mode numbers of 
$Q_1$, $u_1$, $d_1$, etc.
The other columns except the last column show the corresponding 
zero-mode numbers.
The last column shows the total number of zero-modes 
in each row.
In the tables, negative numbers mean matter fields with 
conjugate representations.
Note that in our analysis we do not distinguish 
$L_i$, $H^d_i$ and conjugates of $H^u_i$.

We consider the type VII in eq.(\ref{eq:7-types}).
We choose 
\begin{eqnarray}\label{eq:type-6-1}
(m^{Q1}_{(1)},m^{Q1}_{(2)},m^{Q1}_{(3)})=(3,1,1).
\end{eqnarray}
Other magnetic fluxes $m^{Qi}_{(d)}$ for 
$i=2,3,4,5$ must have vanishing elements for one of $d=1,2,3$.
Various combinations are possible 
\begin{eqnarray}
 & & (m^{Q2}_{(1)},m^{Q3}_{(1)},m^{Q4}_{(1)}) = (0,0,0), \nonumber  \\
 & & (m^{Q2}_{(1)},m^{Q3}_{(1)},m^{Q4}_{(2)}) = (0,0,0), \nonumber  \\
 & & (m^{Q2}_{(1)},m^{Q3}_{(2)},m^{Q4}_{(2)}) = (0,0,0), \nonumber  \\
 & & (m^{Q2}_{(1)},m^{Q3}_{(2)},m^{Q4}_{(3)}) = (0,0,0), \\
 & & (m^{Q2}_{(2)},m^{Q3}_{(2)},m^{Q4}_{(2)}) = (0,0,0), \nonumber  \\
 & & (m^{Q2}_{(2)},m^{Q3}_{(2)},m^{Q4}_{(3)}) = (0,0,0). \nonumber  
\end{eqnarray}
Among them, all the combinations of magnetic fluxes 
leading to the three-generation models are shown in Tables 
26, 28, 30, 32 34, 36 and 38 as the models VII-1,$\cdots$, 8, 
where $n$ denotes arbitrary integer.
Thus, this type includes many semi-realistic models.
The corresponding massless spectra are shown in Tables 
27, 29, 31, 33, 35, 37 and 39.

We have classified the three-generation models with 
the required aspect.
All of the models shown in Tables 2-39 have three chiral 
generations of quarks and leptons, 
several vector-like generations 
and many singlets, 
but matter fields with exotic representations such as 
$Q_Y$ do not appear.

\subsection{Yukawa couplings}

In section 3.2, we have obtained various semi-realistic models.
Here, we study their Yukawa couplings.
Our models have the gauge group 
$SU(3) \times SU(2) \times U(1)_Y \times U(1)^4$.
The top Yukawa coupling must be allowed by 
$SU(3) \times SU(2) \times U(1)_Y \times U(1)^4$.
Most of our models include several singlets with vanishing 
$U(1)_Y$ charge.
Their vacuum expectation values (VEVs) would break extra $U(1)^4$ 
symmetries.
Higher order couplings would lead to 
effective Yukawa couplings through such breaking and 
such effective Yukawa couplings may be small.
Thus, we require that the top Yukawa coupling 
must appear as a 3-point coupling allowed by the 
$SU(3) \times SU(2) \times U(1)_Y \times U(1)^4$.

In section 3.2, the semi-realistic massless spectra
are obtained from the type VI and type VII.
However, any of the models in the 
type VI shown in Tables 2-25 do not allow the top Yukawa 
coupling.
The top Yukawa coupling is allowed 
in only the 
models VII-3, 6 and 8 shown in Tables 29, 35 and 39.
Hence, these models are more interesting than others.

All of the models VII-3, 6 and 8 have many vector-like generations and 
singlets with vanishing $U(1)_Y$, in addition to the three chiral 
generations.
For example, the model VII-6 has ten and fifteen vector-like 
generations for $u$ and $d$, respectively, 
and other models have more vector-like generations.
We expect that such vector-like generations would gain 
effective mass terms from higher order couplings 
including singlets after the symmetry breaking due to 
VEVs of singlets.
For example, in the model VII-6 
we can show that all of the vector-like generations 
have  3-point and 4-point couplings with singlets, 
which would become mass terms of vector-like generations 
after the symmetry breaking.
Phenomenological aspects of our models, 
e.g. quark/lepton mass matrices, depend on patterns of 
many singlet VEVs, i.e. 
which linear combinations would remain as three chiral generations 
and which higher order couplings would become effective 
Yukawa couplings after the symmetry breaking.

Here, for illustration, let us study 
 quark mass matrices with rather simple assumptions.
First, let us consider the model VII-6.
We assume that three chiral generations of $u$ and $d$ are 
originated from $u_2$ and $d_3$.
The Higgs fields $H^u_3$ ($H^d_2$) have allowed Yukawa couplings 
with $Q_1$ and $u_2$ ($d_3$).
The numbers of zero-modes for the $Q_1$ fields are 
equal to $(3,1,1)$ on $T^2_{(1)}$, $T^2_{(2)}$ and 
$T^2_{(3)}$.
Similarly, the numbers of zero-modes for $u_2$ fields 
equal to $(1,1,3)$ on  $T^2_{(1)}$, $T^2_{(2)}$ and 
$T^2_{(3)}$.
In addition, 
the numbers of zero-modes for $H^u_3$ fields 
equal to $(2,2,2)$ on  $T^2_{(1)}$, $T^2_{(2)}$ and 
$T^2_{(3)}$.
Note that the flavor structure of $Q_1$ 
is determined by the first $T^2_{(1)}$, while 
the flavor structure of $u_2$ is determined by 
the third $T^2_{(3)}$.
Thus, for one of Higgs fields $H^u_3$, the 
up-sector Yukawa coupling matrix is always written as 
\begin{eqnarray}\label{eq:rank-1}
Y_{ij}^u= a_i b_j.
\end{eqnarray}
That is a matrix with the rank one.
Only the third generation can be massive, but 
the other two generations are massless 
for one of VEVs of eight $H^u_3$ fields.
Suppose that all of eight $H^u_3$ fields develop 
their VEVs.
Note that the two zero-modes of $H^u_3$ on 
the second  $T^2_{(2)}$ do not lead to variety 
of the mass matrix, because both 
$Q_1$ and $U_2$ have single zero-modes on $T^2_{(2)}$.
Then, the mass matrix induced from 
the 3-point couplings would be written by 
the following form, 
\begin{eqnarray}
m_{ij}^u =  a_i^{(1)} b_j^{(1)} v^{(1,1)} + 
 a_i^{(2)} b_j^{(1)} v^{(2,1)} +  a_i^{(1)} b_j^{(2)} v^{(1,2)} +
  a_i^{(2)} b_j^{(2)} v^{(2,2)} ,
\end{eqnarray}
where $v^{(k,\ell)}$ for $k,\ell=1,2$ denote the VEVs of $H^2_3$ 
fields and $k$ and $\ell$ correspond to the zero-mode indices 
for $T^2_{(1)}$ and $T^2_{(3)}$.
Note that for each of $v^{(k,\ell)}$ the Yukawa matrix has the 
same form as Eq.~(\ref{eq:rank-1}).
The mass matrix $m_{ij}^u$ can be written as 
\begin{eqnarray}
m_{ij}^u =  v^{(1,1)} \left( a_i^{(1)}+  a_i^{(2)} 
\frac{v^{(2,1)}}{v^{(1,1)}} \right) 
\left( b_j^{(1)} + b_j^{(2)} \frac{v^{(1,2)}}{v^{(1,1)}} \right)
+ a_i^{(2)} b_j^{(2)} \left( v^{(2,2)} -\frac{v^{(2,1)}v^{(1,2)}}{v^{(1,1)}}
\right).
\end{eqnarray}
This mass matrix $m_{ij}^u$ corresponds to the rank-two.
That is, the second and third generations are massive, but the first 
generation is massless.
It is straightforward to derive the ratio between 
the charm and top quark masses, because there are several parameters 
such as VEVs $v^{(k,\ell)}$, the complex structure moduli 
and Wilson lines.

The down-sector mass matrix is also the rank-two matrix, 
when we consider only the 3-point couplings with eight $H^d_2$ 
fields.
Thus, the mass ratios $m_c/m_t$ and 
$m_s/m_b$ as well as the mixing angle $V_{cb}$ can be realized 
by choosing proper values of parameters, but
the masses of the first generation, 
$m_u$ and $m_d$, and the mixing angles, 
$V_{us}$ and $V_{ub}$ are vanishing.
They may be induced by effective Yukawa couplings, 
which are obtained from higher order couplings 
through the symmetry breaking.

As another illustrating example, let us study 
the quark mass matrices in the model VII-8 with 
$n=1$.
This model has three generations of $Q$ fields from 
$Q_1$, which have three-zero modes on the first $T^2_{(1)}$.
In addition, this model has many vector-like generations of 
$u$ and $d$.
There are allowed 3-points couplings including singlets, 
such that all of the vector-like generations of $u$ 
and $d$ gain masses after those singlets develop 
their VEVs.
The low-energy phenomenology depends on mass terms 
of those vector-like generations.
For illustration, we study the quark mass matrices with rather 
simple assumptions, again.
For example, if all of three chiral light generations of 
$u$ are originated from $u_2$, we would have the 
same result as in the previous model, i.e. the 
rank-two mass matrices.
To illustrate another possibility, here we consider the case that 
three light generations of $u$ ($d$) are 
originated from one of $u_2$ ($d_2$), one of $u_3$ ($d_3$) 
and one of $u_4$ ($d_4$).
For example, their zero-modes correspond to the $j=0$ mode 
on each of $T^2_{(d)}$.
The $u_2$, $u_3$ and $u_4$ ($d_2$, $d_3$ and $d_4$) fields  
have different extra $U(1)^4$ charges.
Thus, they are affected by different Wilson lines.
This model also has several Higgs fields, which 
have the allowed 3-point couplings with these 
quarks.
The Higgs fields, $H^u_3$, $H^u_2$ and 
the conjugates of $H^d_1$ are allowed to couple with 
$(Q,u_2)$, $(Q,u_3)$ and $(Q,u_4)$, respectively.
Similarly, the Higgs fields, $H^d_3$, $H^d_2$ and 
the conjugates of $H^u_1$ are allowed to couple with 
$(Q,d_2)$, $(Q,d_3)$ and $(Q,d_4)$, respectively.
Each of these Higgs fields has eight total zero-modes.
To reduce the number of free parameters, 
we consider only one zero-mode
for each of these Higgs fields, 
$H^u_3$, $H^u_2$,  $H^d_3$, $H^d_2$ and  
the conjugates of $H^d_1$ and $H^u_1$, 
e.g. the zero-mode corresponding to 
$j=0$ for each of $T^2_{(1)}$.
Furthermore, for simplicity we choose $\tau_d=i$ and 
assume that all of Higgs VEVs are the same.

The Yukawa couplings are given by Eq.~(\ref{Yukawa}).
Since three generations of $Q$ are originated from 
the first $T^2_{(1)}$, the flavor structure is 
determined almost by the first $T^2_{(1)}$, 
while the other tori contribute to the overall factors.
That is, the mass ratios and mixing angles are determined by 
only the first torus $T^2_{(1)}$.
In particular, the Wilson lines are important.
Recall that $u_2$, $u_3$ and $u_4$ fields 
have different extra $U(1)^4$ charges.
Thus, different Wilson lines $\tilde \zeta^{(d)}$ 
appear in the Yukawa couplings (\ref{Yukawa}) 
corresponding to $(Q,u_2)$,$(Q,u_3)$, $(Q,u_4)$.
The $d_2$ field has the same extra $U(1)^4$ charges as $u_2$, 
but obviously has the $U(1)_Y$ charge different from $u_2$.
Thus, different Wilson lines $\tilde \zeta^{(d)}$  can appear 
in the Yukawa couplings (\ref{Yukawa}) corresponding to 
$(Q,u_2)$ and $(Q,d_2)$.
Similarly, the $d_3$ ($d_4$) field have the same 
extra $U(1)^4$ charges as $u_3$ ($u_4$), 
 but has the $U(1)_Y$ charge different from $u_3$ ($u_4$).
Thus, the Wilson lines appearing 
in the Yukawa couplings for $(Q,d_3)$ and $(Q,d_4)$ 
are not independent of the other Wilson lines.
Hence, there are four free parameters for Wilson lines 
on the first torus $T^2_{(1)}$.
For example, we choose 
\begin{eqnarray}
\tilde \zeta^{(1)} = 0.071 {\rm~~~for~~~} (Q,u_2), \nonumber \\
\tilde \zeta^{(1)} = 0.011 {\rm~~~for~~~} (Q,u_3),  \nonumber\\
\tilde \zeta^{(1)} = -0.021 {\rm~~~for~~~} (Q,u_4),  \\
\tilde \zeta^{(1)} = -0.16 {\rm~~~for~~~} (Q,d_2).  \nonumber 
\end{eqnarray}
Then, we can derive the following values 
\begin{eqnarray}
& & m_t/m_c = 73, \qquad m_c/m_u = 41, \nonumber \\
& & m_b/m_s = 69, \qquad m_s/m_d = 46, \\
& & V_{cb} = 0.034, \qquad V_{us}=0.19, \qquad V_{ub}=0.003. \nonumber
\end{eqnarray}
These mixing angles are realistic and 
mass ratios except $m_c/m_u$ are similar to experimental values up to 
a few factors.
We have assumed all of Higgs VEVs are the same and 
taken $\tau_d=i$.
By varying them, we would obtain more realistic values.
Indeed, the number of free parameters is larger than 
the number of observables.

\subsection{Another attempt for realistic models}

In section 3.2, we have classified the three-generation 
models with the supersymmetric condition (\ref{eq:SUSY-condition}).
If this condition is not satisfied, 
the Fayet-Illiopoulos D-terms, which depend on 
magnetic fluxes and the area ${\cal A}_d$, appears 
along extra $U(1)^4$ directions 
in the terminology of 4D N=1 global supersymmetry.
Most of models have many $SU(3) \times SU(2)$ singlets 
with vanishing $U(1)_Y$ charges.
Such singlets may develop their VEVs such that 
they cancel the Fayet-Illiopoulos D-terms and 
a stable vacuum is realized.
Thus, let us systematically search realistic models 
without imposing the condition (\ref{eq:SUSY-condition}).

Here, we concentrate the three-generation models without 
vector-like generations for $Q$, $u$, $e$, $d$.
That correspond to the following conditions 
\begin{eqnarray}\label{eq:condition-u-2}
m^{ui}  \geq 0, \qquad m^{di}  \geq 0, \qquad m^{ei}  \geq 0, 
\end{eqnarray}
in addition to the conditions (\ref{eq:condition-Q1}), 
(\ref{eq:condition-Q2}) and  (\ref{eq:condition-u-3}).
For $L$, $H^u$ and $H^d$, we require the same condition 
as Eq.~(\ref{eq:condition-L}).
Furthermore, we require the condition (\ref{eq:condition-QY}) and 
$m^{Di}=m^{D^ci}=0$.

Under the above conditions, we can find many models, 
which realize exactly the massless spectrum of 
the minimal supersymmetric standard model (MSSM), up to singlets.
For example, we choose the following magnetic fluxes,
\begin{eqnarray}
(m^{Q1}_{(1)},m^{Q1}_{(2)},m^{Q1}_{(3)}) &=& (0,-1,-1), \nonumber \\
(m^{Q2}_{(1)},m^{Q2}_{(2)},m^{Q2}_{(3)}) &=& (0,-2,-1), \nonumber\\
(m^{Q3}_{(1)},m^{Q3}_{(2)},m^{Q3}_{(3)}) &=& (0,1,0), \\
(m^{Q4}_{(1)},m^{Q4}_{(2)},m^{Q4}_{(3)}) &=& (-1,-1,1),\nonumber \\
(m^{Q5}_{(1)},m^{Q5}_{(2)},m^{Q5}_{(3)}) &=& (1,2,1). \nonumber
\end{eqnarray}
This model has the three generations of quarks and leptons, 
and one pair of Higgs fields as well as many 
singlets.\footnote{This model has four zero-modes
  for $L$, and one of them can be considered as $H^d$.}
Thus, this model would be quite interesting from the 
viewpoint of the massless spectrum.
However, the top Yukawa coupling 
is not allowed in this model.
We can find many similar models, where 
the massless spectrum of the MSSM is realized, but 
unfortunately the top Yukawa coupling is not allowed.

Indeed, we can show that there is no model with the 
allowed top Yukawa coupling under the above condition.
For example, let us consider the model, where 
the top Yukawa coupling appears from the 
coupling among the fields, $Q_1$, $u_2$ and $H^u_3$.
Since at least one zero-mode must appear from $Q_1$, 
the possible magnetic fluxes are classified into 
the following three cases,
\begin{eqnarray}
(m^{Q1}_{(1)},m^{Q1}_{(2)},m^{Q1}_{(3)})=(1,1,1), \qquad (2,1,1), 
\qquad (3,1,1).
\end{eqnarray} 
The gauge invariance requires that $m^{Q1}_{(d)}+m^{u2}_{(d)}+
m^{H^u3}_{(d)}=0$.
We require that both $u_2$ and $H^u_3$ have at least 
one zero-modes.
Then, the possible combinations of magnetic fluxes are 
classified into the following four combinations,
\begin{equation}
\left \{
\begin{array}{l}
(m^{Q1}_{(1)},m^{Q1}_{(2)},m^{Q1}_{(3)}) = (3,1,1),   \\
(m^{u2}_{(1)},m^{u2}_{(2)},m^{u2}_{(3)}) = (-1,-3,1),   \\
(m^{H^u3}_{(1)},m^{H^u3}_{(2)},m^{H^u3}_{(3)}) = (-2,2,-2), 
\end{array}
\right.
\end{equation} 
\begin{equation}
\left \{
\begin{array}{l}
(m^{Q1}_{(1)},m^{Q1}_{(2)},m^{Q1}_{(3)}) = (3,1,1),   \\
(m^{u2}_{(1)},m^{u2}_{(2)},m^{u2}_{(3)}) = (-1,-2,1),   \\
(m^{H^u3}_{(1)},m^{H^u3}_{(2)},m^{H^u3}_{(3)}) = (-2,1,-2), 
\end{array}
\right.
\end{equation} 
\begin{equation}
\left \{
\begin{array}{l}
(m^{Q1}_{(1)},m^{Q1}_{(2)},m^{Q1}_{(3)}) = (2,1,1),   \\
(m^{u2}_{(1)},m^{u2}_{(2)},m^{u2}_{(3)}) = (-1,-3,1),   \\
(m^{H^u3}_{(1)},m^{H^u3}_{(2)},m^{H^u3}_{(3)}) = (-1,2,-2), 
\end{array}
\right.
\end{equation} 
\begin{equation}
\left \{
\begin{array}{l}
(m^{Q1}_{(1)},m^{Q1}_{(2)},m^{Q1}_{(3)}) = (2,1,1),   \\
(m^{u2}_{(1)},m^{u2}_{(2)},m^{u2}_{(3)}) = (-1,-2,1),   \\
(m^{H^u3}_{(1)},m^{H^u3}_{(2)},m^{H^u3}_{(3)}) = (-1,1,-2). 
\end{array}
\right.
\end{equation} 
The magnetic fluxes $m^{Qi}_{(d)}$ are constrained such that 
they realize the above values of $m^{u2}_{(d)}$ and 
$m^{H^u3}_{(d)}$.
Among such constrained combinations of $m^{Qi}_{(d)}$, 
we can not find the above three-generation massless spectrum 
without vector-like generations.

In the above, we have considered the case that the 
top Yukawa coupling is originated from the 
coupling among the fields, $Q_1$, $u_2$ and $H^u_3$.
However, the situation is the same for the other allowed couplings.
Hence, we can not obtain the three generation spectrum without 
vector-like generations in the models with the allowed top Yukawa
coupling.

\section{Conclusion}

We have studied 10D $N=1$ super Yang-Mills $E_8$ theory 
on the $(T^2)^3$ background with magnetic fluxes.
We have classified the models with semi-realistic massless 
spectra, that is, three chiral generations and several 
vector-like generations.
We have obtained various three-generation models, but 
the top Yukawa coupling is forbidden in many of them.

We have obtained various semi-realistic models.
However, many vector-like generations are, in general, 
included in those models.
Although we have concentrated to integer magnetic fluxes, 
it is interesting to extend our analysis to the torus 
compactification with fractional fluxes and non-Abelian 
Wilson lines~\cite{toron,CIM,Alfaro:2006is,Salvatori:2006pb,Abe:2010ii}, 
and the orbifold compactifications~\cite{Abe:2008fi,Abe:2009uz}. 
Since such backgrounds could project out some zero-modes, 
they would lead to more interesting models.

\subsection*{Acknowledgement}

The authors would like to thank K.S.~Choi and Y.~Nakai for useful discussions.
T.~K., M.~M. and H.~O. are supported in part by the Grant-in-Aid for 
Scientific Research No.~20540266, No.~21$\cdot$173 and
No.~21$\cdot$897 from the 
Ministry of Education, Culture, Sports, Science and Technology of Japan.
T.~K. is also supported in part by the Grant-in-Aid for the Global COE 
Program "The Next Generation of Physics, Spun from Universality and 
Emergence" from the Ministry of Education, Culture,Sports, Science and 
Technology of Japan.

\begin{table}[t]
\begin{center}
\footnotesize
\begin{tabular}{|c|cccccc|}
\hline 
& $SU(3) \times SU(2)$ & 
$U(1)_a$ &  $U(1)_b$ & $U(1)_c$ &  $U(1)_d$ & $U(1)_Y$ \\ \hline 
$Q_1$ & $({\bf 3},{\bf 2})$ & 1 & 0 & 1 & -1 & 1 \\
$u_1^c$ & $(\bar {\bf 3},{\bf 1})$ & 1 & 0 & 1 & -1 & -4 \\
$e^c_1$ & $({\bf  1},{\bf 1})$ & 1 & 0 & 1 & -1 & 6 \\
$d_1^c$ & $(\bar {\bf 3},{\bf 1})$ & 1 & 0 & 1 & 3 & 2 \\
$L_1$ & $({\bf 1},{\bf 2})$ & 1 & 0 & 1 & 3 & -3 \\
$\nu_1^c$ & $(\bar {\bf 1},{\bf 1})$ & 1 & 0 & 1 & -5 & 0 \\
$D_1$ & $({\bf 3},{\bf 1})$ & 1 & 0 & -2 & 2 & -2 \\
$H^u_1$ & $({\bf 1},{\bf 2})$ & 1 & 0 & -2 & 2 & 3 \\
$D_1^c$ & $(\bar {\bf 3},{\bf 1})$ & 1 & 0 & -2 & -2 & 2 \\
$H^d_1$ & $({\bf 1},{\bf 2})$ & 1 & 0 & -2 & -2 & -3 \\
$S_1$ & $({\bf 1},{\bf 1})$ & 1 & 0 & 4 & 0 & 0 \\
$Q_2$ & $({\bf 3},{\bf 2})$ & 0 & 1 & 1 & -1 & 1 \\
$u_2^c$ & $(\bar {\bf 3},{\bf 1})$ & 0 & 1 & 1 & -1 & -4 \\
$e^c_2$ & $({\bf  1},{\bf 1})$ & 0 & 1 & 1 & -1 & 6 \\
$d_2^c$ & $(\bar {\bf 3},{\bf 1})$ & 0 & 1 & 1 & 3 & 2 \\
$L_2$ & $({\bf 1},{\bf 2})$ & 0 & 1 & 1 & 3 & -3 \\
$\nu_2^c$ & $(\bar {\bf 1},{\bf 1})$ & 0 & 1 & 1 & -5 & 0 \\
$D_2$ & $({\bf 3},{\bf 1})$ & 0 & 1 & -2 & 2 & -2 \\
$H^u_2$ & $({\bf 1},{\bf 2})$ & 0 & 1 & -2 & 2 & 3 \\
$D_2^c$ & $(\bar {\bf 3},{\bf 1})$ & 0 & 1 & -2 & -2 & 2 \\
$H^d_2$ & $({\bf 1},{\bf 2})$ & 0 & 1 & -2 & -2 & -3 \\
$S_2$ & $({\bf 1},{\bf 1})$ & 0 & 1 & 4 & 0 & 0 \\
$Q_3$ & $({\bf 3},{\bf 2})$ & -1 & -1 & 1 & -1 & 1 \\
$u_3^c$ & $(\bar {\bf 3},{\bf 1})$ & -1 & -1 & 1 & -1 & -4 \\
$e^c_3$ & $({\bf  1},{\bf 1})$ & -1 & -1 & 1 & -1 & 6 \\
$d_3^c$ & $(\bar {\bf 3},{\bf 1})$ & -1 & -1 & 1 & 3 & 2 \\
$L_3$ & $({\bf 1},{\bf 2})$ & -1 & -1 & 1 & 3 & -3 \\
$\nu_3^c$ & $(\bar {\bf 1},{\bf 1})$ & -1 & -1 & 1 & -5 & 0 \\
$D_3$ & $({\bf 3},{\bf 1})$ & -1 & -1 & -2 & 2 & -2 \\
$H^u_3$ & $({\bf 1},{\bf 2})$ & -1 & -1 & -2 & 2 & 3 \\
$D_3^c$ & $(\bar {\bf 3},{\bf 1})$ & -1 & -1 & -2 & -2 & 2 \\
$H^d_3$ & $({\bf 1},{\bf 2})$ & -1 & -1 & -2 & -2 & -3 \\
$S_3$ & $({\bf 1},{\bf 1})$ & -1 & -1 & 4 & 0 & 0 \\
$N_1$ & $({\bf 1},{\bf 1})$ & 2 & 1 & 0 & 0 & 0 \\
$N_2$ & $({\bf 1},{\bf 1})$ & 1 & 2 & 0 & 0 & 0 \\
$N_3$ & $({\bf 1},{\bf 1})$ & 1 & -1 & 0 & 0 & 0 \\
$Q_4$ & $({\bf 3},{\bf 2})$ & 0 & 0 & -3 & -1 & 1 \\
$u_4^c$ & $(\bar {\bf 3},{\bf 1})$ & 0 & 0 & -3 & -1 & -4 \\
$e^c_4$ & $({\bf  1},{\bf 1})$ & 0 & 0 & -3 & -1 & 6 \\
$d_4^c$ & $(\bar {\bf 3},{\bf 1})$ & 0 & 0 & -3 & 3 & 2 \\
$L_4$ & $({\bf 1},{\bf 2})$ & 0 & 0 & -3 & 3 & -3 \\
$\nu_4^c$ & $(\bar {\bf 1},{\bf 1})$ & 0 & 0 & -3 & -5 & 0 \\
$Q_5$ & $({\bf 3},{\bf 2})$ & 0 & 0 & 0 & 4 & 1 \\
$u_5^c$ & $(\bar {\bf 3},{\bf 1})$ & 0 & 0 & 0 & 4 & -4 \\
$e^c_5$ & $({\bf  1},{\bf 1})$ & 0 & 0 & 0 & 4 & 6 \\
$Q_Y$ & $(\bar {\bf 3},{\bf 2})$ & 0 & 0 & 0 & 0 & 5 \\
\hline  
\end{tabular}

\caption{Decomposition of the $E_8$  ${\bf 248}$ adjoint representation 
in $SU(3) \times SU(2) \times U(1)_Y \times U(1)^4$} 
\label{tab:248}
\end{center}
\end{table}

\clearpage

\begin{table}[t]
\begin{center}
\begin{tabular}{c|ccccc|c} $d$ 
& $m^{Q1}_{(d)}$ & $m^{Q2}_{(d)}$ & $m^{Q3}_{(d)}$ & $m^{Q4}_{(d)}$ &
  $m^{Q5}_{(d)}$ & $m^{QY}_{(d)}$ \\
\hline
$1$ & 0 & 0 & 0 & 0 & 3 & 3 \\
$2$ & 0 & 1 & -1 & -1 & 1 & 0   \\
$3$ & 0 & -1 & 1 & 1 & 1 & 2  \\
\hline
\end{tabular}
\caption{Magnetic fluxes in the type VI (model VI-1) }
\end{center}
\end{table}

\vspace{0.2cm}
\begin{table}[t]
\begin{center}
\begin{tabular}{c|c|c|c|c|c|c} $i$ & 1 & 2 & 3 & 4 & 5 & sum \\
\hline
$Q_i$ & $0$ & $0$ & $0$ & $0$ & $3$ & $3$  \\
$u_i$ & $0$ & $9$ & $-3$ & $-3$ & $0$ & $3$ \\
$e_i$ & $0$ & $3$ & $-9$ & $-9$ & $18$ & $3$ \\
$d_i$ & $3$ & $0$ & $0$ & $0$ & & $3$  \\
$L_i$ & $0$ & $0$ & $0$ & $0$ & & $0$ \\
$H^u_i$ & $0$ & $3$ & $-9$ & & & $-6$ \\
$H^d_i$ & $-3$ & $0$ & $0$ & & & $-3$ \\ 
$\nu^c_i$ &-3&0&0&0&& -3\\
$S_i$ &0&0&0&&& 0\\
$N_i$ &0&0&0&&& 0\\
\hline
\end{tabular}
\caption{Massless spectrum in the model VI-1 }
\end{center}
\end{table}

\vspace{0.8cm}
\begin{table}[t]
\begin{center}
\begin{tabular}{c|ccccc|c} $d$ & $m^{Q1}_{(d)}$ & $m^{Q2}_{(d)}$ & $m^{Q3}_{(d)}$ & $m^{Q4}_{(d)}$ & $m^{Q5}_{(d)}$ & $m^{QY}_{(d)}$ \\
\hline
$1$ & -1 & -1 & -1 & 0 & 3 & 0 \\
$2$ & 0 & 0 & 0 & 0 & 1 & 1   \\
$3$ & -1 & -1 & -1 & 0 & 1 & -2  \\
\hline
\end{tabular}
\caption{Magnetic fluxes in the type VI (model VI-2) }
\end{center}
\end{table}

\vspace{0.2cm}
\begin{table}[t]
\begin{center}
\begin{tabular}{c|c|c|c|c|c|c} $i$  & 1 & 2 & 3 & 4 & 5 & sum \\
\hline
$Q_i$ & $0$ & $0$ & $0$ & $0$ & $3$ & $3$  \\
$u_i$ & $1$ & $1$ & $1$ & $0$ & $0$ & $3$ \\
$e_i$ & $3$ & $3$ & $3$ & $0$ & $-6$ & $3$ \\
$d_i$ & $0$ & $0$ & $0$ & $3$ & & $3$  \\
$L_i$ & $0$ & $0$ & $0$ & $0$ & & $0$ \\
$H^u_i$ & $0$ & $0$ & $0$ & & & $0$ \\
$H^d_i$ & $1$ & $1$ & $1$ & & & $3$ \\
$\nu^c_i$ &-8&-8&-8&-3&& -27\\
$S_i$ &0&0&0&&& 0\\
$N_i$ &0&0&0&&& 0\\ 
\hline
\end{tabular}
\caption{Massless spectrum in the model VI-2 }
\end{center}
\end{table}

\vspace{0.8cm}
\begin{table}[t]
\begin{center}
\begin{tabular}{c|ccccc|c} $d$ & $m^{Q1}_{(d)}$ & $m^{Q2}_{(d)}$ & $m^{Q3}_{(d)}$ & $m^{Q4}_{(d)}$ & $m^{Q5}_{(d)}$ & $m^{QY}_{(d)}$ \\
\hline
$1$ & -3 & 2 & -1 & -1 & 3 & 0 \\
$2$ & 0 & 0 & 0 & 0 & 1 & 1   \\
$3$ & 3 & -2 & 1 & 1 & 1 & 4  \\
\hline
\end{tabular}
\caption{Magnetic fluxes in the type VI (model VI-3) }
\end{center}
\end{table}

\vspace{0.1cm}
\begin{table}[t]
\begin{center}
\begin{tabular}{c|c|c|c|c|c|c} $i$ & 1 & 2 & 3 & 4 & 5 & sum \\
\hline
$Q_i$ & $0$ & $0$ & $0$ & $0$ & $3$ & $3$  \\
$u_i$ & $-3$ & $12$ & $-3$ & $-3$ & $0$ & $3$ \\
$e_i$ & $-21$ & $4$ & $-5$ & $-5$ & $30$ & $3$ \\
$d_i$ & $0$ & $-5$ & $4$ & $4$ & & $3$  \\
$L_i$ & $0$ & $0$ & $0$ & $0$ & & $0$ \\
$H^u_i$ & $-5$ & $0$ & $3$ & & & $-2$ \\
$H^d_i$ & $0$ & $5$ & $-4$ & & & $1$ \\
$\nu^c_i$ &12&-3&0&0&& 9\\
$S_i$ &0&0&0&&& 0\\
$N_i$ &0&0&0&&& 0\\ 
\hline
\end{tabular}
\caption{Massless spectrum in the model VI-3 }
\end{center}
\end{table}

\vspace{0.5cm}
\begin{table}[t]
\begin{center}
\begin{tabular}{c|ccccc|c}  $d$ & $m^{Q1}_{(d)}$ & $m^{Q2}_{(d)}$ & $m^{Q3}_{(d)}$ & $m^{Q4}_{(d)}$ & $m^{Q5}_{(d)}$ & $m^{QY}_{(d)}$ \\
\hline
$1$ & -3 & -2 & 1 & 1 & 3 & 0 \\
$2$ & 0 & 0 & 0 & 0 & 1 & 1   \\
$3$ & 3 & 2 & -1 & -1 & 1 & 4  \\
\hline
\end{tabular}
\caption{Magnetic fluxes in the type VI (model VI-4)}
\end{center}
\end{table}

\vspace{0.1cm}
\begin{table}[t]
\begin{center}
\begin{tabular}{c|c|c|c|c|c|c} $i$ & 1 & 2 & 3 & 4 & 5 & sum \\
\hline
$Q_i$ & $0$ & $0$ & $0$ & $0$ & $3$ & $3$  \\
$u_i$ & $-3$ & $-4$ & $5$ & $5$ & $0$ & $3$ \\
$e_i$ & $-21$ & $-12$ & $3$ & $3$ & $30$ & $3$ \\
$d_i$ & $0$ & $3$ & $0$ & $0$ & & $3$  \\
$L_i$ & $0$ & $0$ & $0$ & $0$ & & $0$ \\
$H^u_i$ & $3$ & $4$ & $-5$ & & & $2$ \\
$H^d_i$ & $-4$ & $-3$ & $12$ & & & $5$ \\
$\nu^c_i$ &12&5&-4&-4&& 9\\
$S_i$ &0&0&0&&& 0 \\
$N_i$ &0&0&0&&& 0\\ 
\hline
\end{tabular}
\caption{Massless spectrum in the model VI-4 }
\end{center}
\end{table}

\clearpage

\vspace{0.5cm}
\begin{table}[t]
\begin{center}
\begin{tabular}{c|ccccc|c} $d$ & $m^{Q1}_{(d)}$ & $m^{Q2}_{(d)}$ & $m^{Q3}_{(d)}$ & $m^{Q4}_{(d)}$ & $m^{Q5}_{(d)}$ & $m^{QY}_{(d)}$ \\
\hline
$1$ & -2 & -1 & -1 & 1 & 3 & 0 \\
$2$ & 0 & 0 & 0 & 0 & 1 & 1   \\
$3$ & -6 & -3 & -3 & 3 & 1 & -8  \\
\hline
\end{tabular}
\caption{Magnetic fluxes in the type VI (model VI-5) }
\end{center}
\end{table}

\vspace{0.1cm}
\begin{table}[t]
\begin{center}
\begin{tabular}{c|c|c|c|c|c|c} $i$ & 1 & 2 & 3 & 4 & 5 & sum \\
\hline
$Q_i$ & $0$ & $0$ & $0$ & $0$ & $3$ & $3$  \\
$u_i$ & $4$ & $5$ & $5$ & $-11$ & $0$ & $3$ \\
$e_i$ & $28$ & $11$ & $11$ & $-5$ & $-42$ & $3$ \\
$d_i$ & $-5$ & $-4$ & $-4$ & $16$ & & $3$  \\
$L_i$ & $0$ & $0$ & $0$ & $0$ & & $0$ \\
$H^u_i$ & $-4$ & $3$ & $3$ & & & $2$ \\
$H^d_i$ & $5$ & $0$ & $0$ & & & $5$ \\
$\nu^c_i$ &-35&-16&-16&4&& -63\\
$S_i$ &0&0&0&&& 0\\
$N_i$ &0&0&0&&& 0\\ 
\hline
\end{tabular}
\caption{Massless spectrum in the model VI-5 }
\end{center}
\end{table}

\vspace{0.5cm}
\begin{table}[t]
\begin{center}
\begin{tabular}{c|ccccc|c} $d$  & $m^{Q1}_{(d)}$ & $m^{Q2}_{(d)}$ & $m^{Q3}_{(d)}$ & $m^{Q4}_{(d)}$ & $m^{Q5}_{(d)}$ & $m^{QY}_{(d)}$ \\
\hline
$1$ & -3 & -1 & 1 & 0 & 3 & 0 \\
$2$ & 0 & 0 & 0 & 0 & 1 & 1   \\
$3$ & 9 & 3 & -3 & 0 & 1 & 10  \\
\hline
\end{tabular}
\caption{Magnetic fluxes in the type VI (model VI-6) }
\end{center}
\end{table}

\vspace{0.1cm}
\begin{table}[t]
\begin{center}
\begin{tabular}{c|c|c|c|c|c|c} $i$ & 1 & 2 & 3 & 4 & 5 & sum \\
\hline
$Q_i$ & $0$ & $0$ & $0$ & $0$ & $3$ & $3$  \\
$u_i$ & $-3$ & $-7$ & $13$ & $0$ & $0$ & $3$ \\
$e_i$ & $-57$ & $-13$ & $7$ & $0$ & $66$ & $3$ \\
$d_i$ & $0$ & $8$ & $-8$ & $3$ & & $3$  \\
$L_i$ & $0$ & $0$ & $0$ & $0$ & & $0$ \\
$H^u_i$ & $0$ & $8$ & $-8$ & & & $0$ \\
$H^d_i$ & $-3$ & $-7$ & $13$ & & & $3$ \\
$\nu^c_i$ &48&8&-8&-3&& 45\\
$S_i$ &0&0&0&&& 0\\
$N_i$ &0&0&0&&& 0\\ 
\hline
\end{tabular}
\caption{Massless spectrum in the model VI-6 }
\end{center}
\end{table}

\vspace{0.5cm}
\begin{table}[t]
\begin{center}
\begin{tabular}{c|ccccc|c} $d$ & $m^{Q1}_{(d)}$ & $m^{Q2}_{(d)}$ & $m^{Q3}_{(d)}$ & $m^{Q4}_{(d)}$ & $m^{Q5}_{(d)}$ & $m^{QY}_{(d)}$ \\
\hline
$1$ & 3 & -3 & -3 & 0 & 3 & 0 \\
$2$ & 0 & 0 & 0 & 0 & 1 & 1   \\
$3$ & -1 & 1 & 1 & 0 & 1 & 2  \\
\hline
\end{tabular}
\caption{Magnetic fluxes in the type VI (model VI-7)}
\end{center}
\end{table}

\vspace{0.1cm}
\begin{table}[t]
\begin{center}
\begin{tabular}{c|c|c|c|c|c|c} $i$ & 1 & 2 & 3 & 4 & 5 & sum \\
\hline
$Q_i$ & $0$ & $0$ & $0$ & $0$ & $3$ & $3$  \\
$u_i$ & $9$ & $-3$ & $-3$ & $0$ & $0$ & $3$ \\
$e_i$ & $3$ & $-9$ & $-9$ & $0$ & $18$ & $3$ \\
$d_i$ & $0$ & $0$ & $0$ & $3$ & & $3$  \\
$L_i$ & $0$ & $0$ & $0$ & $0$ & & $0$ \\
$H^u_i$ & $0$ & $0$ & $0$ & & & $0$ \\
$H^d_i$ & $9$ & $-3$ & $-3$ & & & $3$ \\
$\nu^c_i$ &0&0&0&-3&& -3\\
$S_i$ &0&0&0&&& 0\\
$N_i$ &0&0&0&&& 0\\ 
\hline
\end{tabular}
\caption{Massless spectrum in the model VI-7 }
\end{center}
\end{table}

\vspace{0.5cm}
\begin{table}[b]
\begin{center}
\begin{tabular}{c|ccccc|c} $d$  & $m^{Q1}_{(d)}$ & $m^{Q2}_{(d)}$ & $m^{Q3}_{(d)}$ & $m^{Q4}_{(d)}$ & $m^{Q5}_{(d)}$ & $m^{QY}_{(d)}$ \\
\hline
$1$ & -2 & 1 & 1 & 1 & 3 & 4 \\
$2$ & 0 & 0 & 0 & 0 & 1 & 1   \\
$3$ & 2 & -1 & -1 & -1 & 1 & 0  \\
\hline
\end{tabular}
\caption{Magnetic fluxes in the type VI (model VI-8)}
\end{center}
\end{table}

\vspace{0.1cm}
\begin{table}[t]
\begin{center}
\begin{tabular}{c|c|c|c|c|c|c} $i$ & 1 & 2 & 3 & 4 & 5 & sum \\
\hline
$Q_i$ & $0$ & $0$ & $0$ & $0$ & $3$ & $3$  \\
$u_i$ & $12$ & $-3$ & $-3$ & $-3$ & $0$ & $3$ \\
$e_i$ & $4$ & $-5$ & $-5$ & $-5$ & $14$ & $3$ \\
$d_i$ & $3$ & $0$ & $0$ & $0$ & & $3$  \\
$L_i$ & $0$ & $0$ & $0$ & $0$ & & $0$ \\
$H^u_i$ & $4$ & $-5$ & $-5$ & & & $-6$ \\
$H^d_i$ & $5$ & $-4$ & $-4$ & & & $-3$ \\
$\nu^c_i$ &5&-4&-4&-4&& -7\\
$S_i$ &0&0&0&&& 0\\
$N_i$ &0&0&0&&& 0\\ 
\hline
\end{tabular}
\caption{Massless spectrum in the model VI-8 }
\end{center}
\end{table}

\vspace{0.5cm}
\begin{table}[t]
\begin{center}
\begin{tabular}{c|ccccc|c} $d$ & $m^{Q1}_{(d)}$ & $m^{Q2}_{(d)}$ & $m^{Q3}_{(d)}$ & $m^{Q4}_{(d)}$ & $m^{Q5}_{(d)}$ & $m^{QY}_{(d)}$ \\
\hline
$1$ & 3 & 3 & -3 & 0 & 3 & 6 \\
$2$ & 0 & 0 & 0 & 0 & 1 & 1   \\
$3$ & -1 & -1 & 1 & 0 & 1 & 0  \\
\hline
\end{tabular}
\caption{Magnetic fluxes in the type VI (model VI-9) }
\end{center}
\end{table}

\vspace{0.1cm}
\begin{table}[t]
\begin{center}
\begin{tabular}{c|c|c|c|c|c|c} $i$ & 1 & 2 & 3 & 4 & 5 & sum \\
\hline
$Q_i$ & $0$ & $0$ & $0$ & $0$ & $3$ & $3$  \\
$u_i$ & $-3$ & $-3$ & $9$ & $0$ & $0$ & $3$ \\
$e_i$ & $-9$ & $-9$ & $3$ & $0$ & $18$ & $3$ \\
$d_i$ & $0$ & $0$ & $0$ & $3$ & & $3$  \\
$L_i$ & $0$ & $0$ & $0$ & $0$ & & $0$ \\
$H^u_i$ & $0$ & $0$ & $0$ & & & $0$ \\
$H^d_i$ & $-3$ & $-3$ & $9$ & & & $3$ \\
$\nu^c_i$ &0&0&0&-3&& -3\\
$S_i$ &0&0&0&&& 0\\
$N_i$ &0&0&0&&& 0\\ 
\hline
\end{tabular}
\caption{Massless spectrum in the model VI-9 }
\end{center}
\end{table}

\vspace{0.5cm}
\begin{table}[t]
\begin{center}
\begin{tabular}{c|ccccc|c} $d$  & $m^{Q1}_{(d)}$ & $m^{Q2}_{(d)}$ & $m^{Q3}_{(d)}$ & $m^{Q4}_{(d)}$ & $m^{Q5}_{(d)}$ & $m^{QY}_{(d)}$ \\
\hline
$1$ & 2 & 1 & -1 & -1 & 3 & 4 \\
$2$ & 0 & 0 & 0 & 0 & 1 & 1   \\
$3$ & -2 & -1 & 1 & 1 & 1 & 0  \\
\hline
\end{tabular}
\caption{Magnetic fluxes in the type VI (model VI-10) }
\end{center}
\end{table}

\begin{table}[t]
\begin{center}
\begin{tabular}{c|c|c|c|c|c|c} $i$ & 1 & 2 & 3 & 4 & 5 & sum \\
\hline
$Q_i$ & $0$ & $0$ & $0$ & $0$ & $3$ & $3$  \\
$u_i$ & $-4$ & $-3$ & $5$ & $5$ & $0$ & $3$ \\
$e_i$ & $-12$ & $-5$ & $3$ & $3$ & $14$ & $3$ \\
$d_i$ & $-5$ & $0$ & $4$ & $4$ & & $3$  \\
$L_i$ & $0$ & $0$ & $0$ & $0$ & & $0$ \\
$H^u_i$ & $0$ & $3$ & $3$ & & & $6$ \\
$H^d_i$ & $-3$ & $0$ & $12$ & & & $9$ \\
$\nu^c_i$ &-3&-4&0&0&& -7\\
$S_i$ &0&0&0&&& 0\\
$N_i$ &0&0&0&&& 0\\ 
\hline
\end{tabular}
\caption{Massless spectrum in the model VI-10 }
\end{center}
\end{table}

\clearpage

\vspace{0.3cm}
\begin{table}[t]
\begin{center}
\begin{tabular}{c|ccccc|c} $d$ & $m^{Q1}_{(d)}$ & $m^{Q2}_{(d)}$ & $m^{Q3}_{(d)}$ & $m^{Q4}_{(d)}$ & $m^{Q5}_{(d)}$ & $m^{QY}_{(d)}$ \\
\hline
$1$ & 3 & 0 & 0 & 0 & 3n & 3+3n \\
$2$ & 1 & -1 & -1 & 1 & 0 & 0   \\
$3$ & 1 & 1 & 1 & -1 & 2n & 2+2n  \\
\hline
\end{tabular}
\caption{Magnetic fluxes in the type VII (model VII-1)}
\end{center}
\end{table}

\vspace{0.1cm}
\begin{table}[t]
\begin{center}
\begin{tabular}{c|c|c|c|c|c|c} $i$ & 1 & 2 & 3 & 4 & 5 & sum \\
\hline
$Q_i$ & $3$ & $0$ & $0$ & $0$ & $0$ & $3$  \\
$u_i$ & $6n^2+3n$ & $-6n^2-9n-3$ & $-6n^2-9n-3$ & $6n^2+15n+9$ & $0$ & $3$ \\
$e_i$ & $6n^2+21n+18$ & $-6n^2-15n-9$ & $-6n^2-15n-9$ & $6n^2+9n+3$ & $0$ & $3$ \\
$d_i$ & $6n^2+9n+3$ & $-6n^2-3n$ & $-6n^2-3n$ & $6n^2-3n$ & & 3  \\
$L_i$ & $0$ & $-3$ & $-3$ & $9$ & & $3$ \\
$H^u_i$ & $12n^2+12n$ & $0$ & $0$ & & & $12n^2+12n$ \\
$H^d_i$ & $12n^2+12n$ & $0$ & $0$ & & & $12n^2+12n$ \\
$\nu^c_i$ &$6n^2-9n+3$&$-6n^2+3n$&$-6n^2+3n$&$6n^2+3n$&& 3\\
$S_i$ &0&0&0&&& 0\\
$N_i$ &0&0&0&&& 0\\ 
\hline
\end{tabular}
\caption{Massless spectrum in the model VII-1 }
\end{center}
\end{table}

\clearpage

\vspace{0.5cm}
\begin{table}[t]
\begin{center}
\begin{tabular}{c|ccccc|c} $d$ & $m^{Q1}_{(d)}$ & $m^{Q2}_{(d)}$ & $m^{Q3}_{(d)}$ & $m^{Q4}_{(d)}$ & $m^{Q5}_{(d)}$ & $m^{QY}_{(d)}$ \\
\hline
$1$ & 3 & -3 & -3 & 3 & 0 & 0 \\
$2$ & 1 & 0 & 0 & 0 & n & 1+n   \\
$3$ & 1 & 1 & 1 & -1 & 2n & 2+2n  \\
\hline
\end{tabular}
\caption{Magnetic fluxes in the type VII (model VII-2)}
\end{center}
\end{table}

\vspace{0.1cm}
\begin{table}[t]
\begin{center}
\begin{tabular}{c|c|c|c|c|c|c} $i$  & 1 & 2 & 3 & 4 & 5 & sum \\
\hline
$Q_i$ & $3$ & $0$ & $0$ & $0$ & $0$ & $3$  \\
$u_i$ & $6n^2+3n$ & $-6n^2-9n-3$ & $-6n^2-9n-3$ & $6n^2+15n+9$ & 0 & 3 \\
$e_i$ & $6n^2+21n+18$ & $-6n^2-15n-9$ & $-6n^2-15n-9$ & $6n^2+9n+3$ & 0 & 3 \\
$d_i$ & $6n^2+9n+3$ & $-6n^2-3n$ & $-6n^2-3n$ & $6n^2-3n$ & & 3  \\
$L_i$ & $0$ & $-3$ & $-3$ & $9$ & & $3$ \\
$H^u_i$ & $12n^2+12n$ & $0$ & $0$ & & & $12n^2+12n$ \\
$H^d_i$ & $12n^2+12n$ & $0$ & $0$ & & & $12n^2+12n$ \\
$\nu^c_i$ &$6n^2-9n+3$&$-6n^2+3n$&$-6n^2+3n$&$6n^2+3n$&& 3\\
$S_i$ &0&0&0&&& 0\\
$N_i$ &0&0&0&&& 0\\ 
\hline
\end{tabular}
\caption{Massless spectrum in the model VII-2 }
\end{center}
\end{table}

\vspace{0.5cm}
\begin{table}[t]
\begin{center}
\begin{tabular}{c|ccccc|c} $d$  & $m^{Q1}_{(d)}$ & $m^{Q2}_{(d)}$ & $m^{Q3}_{(d)}$ & $m^{Q4}_{(d)}$ & $m^{Q5}_{(d)}$ & $m^{QY}_{(d)}$ \\
\hline
$1$ & 3 & -3 & -3 & -1 & 4 & 0 \\
$2$ & 1 & 0 & 0 & 0 & 2 & 3   \\
$3$ & 1 & -3 & -3 & -1 & 0 & -6  \\
\hline
\end{tabular}
\caption{Magnetic fluxes in the type VII (model VII-3)}
\end{center}
\end{table}

\vspace{0.1cm}
\begin{table}[t]
\begin{center}
\begin{tabular}{c|c|c|c|c|c|c} $i$ & 1 & 2 & 3 & 4 & 5 & sum \\
\hline
$Q_i$ & $3$ & $0$ & $0$ & $0$ & $0$ & $3$  \\
$u_i$ & $-42$ & $27$ & $27$ & $15$ & $-24$ & $3$ \\
$e_i$ & $-60$ & $81$ & $81$ & $21$ & $-120$ & $3$ \\
$d_i$ & $21$ & $-6$ & $-6$ & $-6$ & & 3  \\
$L_i$ & $0$ & $-3$ & $-3$ & $-15$ & & $-21$ \\
$H^u_i$ & $0$ & $0$ & $0$ & & & $0$ \\
$H^d_i$ & $-24$ & $24$ & $24$ & & & $24$ \\
$\nu^c_i$ &1&-42&-42&-10&& -93\\
$S_i$ &8&0&0&&& 8\\
$N_i$ &24&0&24&&& 48\\ 
\hline
\end{tabular}
\caption{Massless spectrum in the model VII-3}
\end{center}
\end{table}

\begin{table}[t]
\begin{center}
\begin{tabular}{c|ccccc|c} $d$ & $m^{Q1}_{(d)}$ & $m^{Q2}_{(d)}$ & $m^{Q3}_{(d)}$ & $m^{Q4}_{(d)}$ & $m^{Q5}_{(d)}$ & $m^{QY}_{(d)}$ \\
\hline
$1$ & 3 & -3 & 3 & 3 & 6n & 6n+6 \\
$2$ & 1 & 0 & 0 & 0 & n & 1+n   \\
$3$ & 1 & 1 & -1 & -1 & 0 & 0  \\
\hline
\end{tabular}
\caption{Magnetic fluxes in the type VII (model VII-4)}
\end{center}
\end{table}

\vspace{0.1cm}
\begin{table}[t]
\begin{center}
\begin{tabular}{c|c|c|c|c|c|c} $i$  & 1 & 2 & 3 & 4 & 5 & sum \\
\hline
$Q_i$ & $3$ & $0$ & $0$ & $0$ & $0$ & $3$  \\
$u_i$ & $6n^2+3n$ & $6n^2+15n+9$ & $-6n^2-9n-3$ & $-6n^2-9n-3$ & 0 & 3 \\
$e_i$ & $6n^2+21n+18$ & $6n^2+9n+3$ & $-6n^2-15n-9$ & $-6n^2-15n-9$ & 0 & 3 \\
$d_i$ & $6n^2+9n+3$ & $6n^2-3n$ & $-6n^2-3n$ & $-6n^2-3n$ & & 3  \\
$L_i$ & $0$ & $9$ & $-3$ & $-3$ & & $3$ \\
$H^u_i$ & $0$ & $0$ & $-12n^2-12n$ & & & $-12n^2-12n$ \\
$H^d_i$ & $0$ & $0$ & $-12n^2-12n$ & & & $-12n^2-12n$ \\
$\nu^c_i$ &$6n^2-9n+3$&$6n^2+3n$&$6n^2+3n$&$-6n^2+3n$&& 3\\
$S_i$ &0&0&0&&& 0\\
$N_i$ &0&0&0&&& 0\\ 
\hline
\end{tabular}
\caption{Massless spectrum in the model VII-4}
\end{center}
\end{table}

\vspace{0.5cm}
\begin{table}[t]
\begin{center}
\begin{tabular}{c|ccccc|c} $d$ & $m^{Q1}_{(d)}$ & $m^{Q2}_{(d)}$ & $m^{Q3}_{(d)}$ & $m^{Q4}_{(d)}$ & $m^{Q5}_{(d)}$ & $m^{QY}_{(d)}$ \\
\hline
$1$ & 3 & -3 & 1 & 1 & -2 & 0 \\
$2$ & 1 & 0 & 0 & 0 & 0 & 1   \\
$3$ & 1 & 3 & -1 & -1 & 2 & 4  \\
\hline
\end{tabular}
\caption{Magnetic fluxes in the type VII (model VII-5)}
\end{center}
\end{table}

\vspace{0.1cm}
\begin{table}[t]
\begin{center}
\begin{tabular}{c|c|c|c|c|c|c} $i$  & 1 & 2 & 3 & 4 & 5 & sum \\
\hline
$Q_i$ & $3$ & $0$ & $0$ & $0$ & $0$ & $3$  \\
$u_i$ & $0$ & $-3$ & $5$ & $5$ & $-4$ & $3$ \\
$e_i$ & $30$ & $-21$ & $3$ & $3$ & $-12$ & $3$ \\
$d_i$ & $3$ & $0$ & $0$ & $0$ & & $3$  \\
$L_i$ & $0$ & $5$ & $-3$ & $-3$ & & $-1$ \\
$H^u_i$ & $4$ & $0$ & $0$ & & & $4$ \\
$H^d_i$ & $0$ & $-4$ & $12$ & & & $8$ \\
$\nu^c_i$ &-5&0&0&0&& -5\\
$S_i$ &4&0&0&&& 4\\
$N_i$ &4&0&-12&&& -8\\ 
\hline
\end{tabular}
\caption{Massless spectrum in the model VII-5 }
\end{center}
\end{table}

\vspace{0.5cm}
\begin{table}[t]
\begin{center}
\begin{tabular}{c|ccccc|c} $d$  & $m^{Q1}_{(d)}$ & $m^{Q2}_{(d)}$ & $m^{Q3}_{(d)}$ & $m^{Q4}_{(d)}$ & $m^{Q5}_{(d)}$ & $m^{QY}_{(d)}$ \\
\hline
$1$ & 3 & -3 & -1 & -1 & 0 & -2 \\
$2$ & 1 & 0 & 0 & 0 & -2 & -1   \\
$3$ & 1 & -3 & -1 & -1 & 4 & 0  \\
\hline
\end{tabular}
\caption{Magnetic fluxes in the type VII (model VII-6) }
\end{center}
\end{table}

\vspace{0.1cm}
\begin{table}[t]
\begin{center}
\begin{tabular}{c|c|c|c|c|c|c} $i$ & 1 & 2 & 3 & 4 & 5 & sum \\
\hline
$Q_i$ & $3$ & $0$ & $0$ & $0$ & $0$ & $3$  \\
$u_i$ & $10$ & $3$ & $-1$ & $-1$ & $-8$ & $3$ \\
$e_i$ & $0$ & $-15$ & $-3$ & $-3$ & $24$ & $3$ \\
$d_i$ & $-15$ & $6$ & $6$ & $6$ & & $3$  \\
$L_i$ & $0$ & $1$ & $-3$ & $-3$ & & $-5$ \\
$H^u_i$ & $-8$ & $0$ & $8$ & & & $0$ \\
$H^d_i$ & $0$ & $8$ & $0$ & & & $8$ \\
$\nu^c_i$ &-27&42&10&10&& 35\\
$S_i$ &8&0&0&&& 8\\
$N_i$ &8&0&24&&& 32\\ 
\hline
\end{tabular}
\caption{Massless spectrum in the model VII-6 }
\end{center}
\end{table}

\vspace{0.5cm}
\begin{table}[t]
\begin{center}
\begin{tabular}{c|ccccc|c} $d$ & $m^{Q1}_{(d)}$ & $m^{Q2}_{(d)}$ & $m^{Q3}_{(d)}$ & $m^{Q4}_{(d)}$ & $m^{Q5}_{(d)}$ & $m^{QY}_{(d)}$ \\
\hline
$1$ & 3 & 1 & 1 & 1 & -2 & 4 \\
$2$ & 1 & 0 & 0 & 0 & 0 & 1   \\
$3$ & 1 & -1 & -1 & -1 & 2 & 0  \\
\hline
\end{tabular}
\caption{Magnetic fluxes in the type VII (model VII-7) }
\end{center}
\end{table}

\vspace{0.1cm}
\begin{table}[t]
\begin{center}
\begin{tabular}{c|c|c|c|c|c|c} $i$ & 1 & 2 & 3 & 4 & 5 & sum \\
\hline
$Q_i$ & $3$ & $0$ & $0$ & $0$ & $0$ & $3$  \\
$u_i$ & $0$ & $-3$ & $-3$ & $-3$ & $12$ & $3$ \\
$e_i$ & $14$ & $-5$ & $-5$ & $-5$ & $4$ & $3$ \\
$d_i$ & $3$ & $0$ & $0$ & $0$ & & $3$  \\
$L_i$ & $0$ & $5$ & $5$ & $5$ & & $15$ \\
$H^u_i$ & $4$ & $0$ & $0$ & & & $4$ \\
$H^d_i$ & $0$ & $-4$ & $-4$ & & & $-8$ \\
$\nu^c_i$ &-5&0&0&0&& -5\\
$S_i$ &4&0&0&&& 4\\
$N_i$ &4&0&4&&& 8\\ 
\hline
\end{tabular}
\caption{Massless spectrum in the model VII-7}
\end{center}
\end{table}

\vspace{0.5cm}
\begin{table}[t]
\begin{center}
\begin{tabular}{c|ccccc|c} $d$  & $m^{Q1}_{(d)}$ & $m^{Q2}_{(d)}$ & $m^{Q3}_{(d)}$ & $m^{Q4}_{(d)}$ & $m^{Q5}_{(d)}$ & $m^{QY}_{(d)}$ \\
\hline
$1$ & 3 & -1 & -1 & -1 & 0 & 0 \\
$2$ & 1 & 0 & 0 & 0 & n & 1+n   \\
$3$ & 1 & -1 & -1 & -1 & -2n & -2-2n  \\
\hline
\end{tabular}
\caption{Magnetic fluxes in the type VII (model VII-8)}
\end{center}
\end{table}

\vspace{0.1cm}
\begin{table}[t]
\begin{center}
\begin{tabular}{c|c|c|c|c|c|c} $i$ & 1 & 2 & 3 & 4 & 5 & sum \\
\hline
$Q_i$ & $3$ & $0$ & $0$ & $0$ & $0$ & $3$  \\
$u_i$ & $-6n^2-9n$ & $2n^2+3n+1$ & $2n^2+3n+1$ & $2n^2+3n+1$ & $0$ & $3$ \\
$e_i$ & $-6n^2-15n-6$ & $2n^2+5n+3$ & $2n^2+5n+3$ & $2n^2+5n+3$ & $0$ & $3$ \\
$d_i$ & $-6n^2-3n+3$ & $2n^2+n$ & $2n^2+n$ & $2n^2+n$ & & $3$  \\
$L_i$ & $0$ & $1$ & $1$ & $1$ & & $3$ \\
$H^u_i$ & $-4n^2-4n$ & $4n^2+4n$ & $4n^2+4n$ & & & $4n^2+4n$ \\
$H^d_i$ & $-4n^2-4n$ & $4n^2+4n$ & $4n^2+4n$ & & & $4n^2+4n$ \\
$\nu^c_i$ &$-6n^2+3n+3$&$2n^2-n$&$2n^2-n$&$2n^2-n$&& 3\\
$S_i$ &8&0&0&&& 8\\
$N_i$ &8&0&8&&& 16\\ 
\hline
\end{tabular}
\caption{Massless spectrum in the model VII-8 }
\end{center}
\end{table}


\begin{thebibliography}{99}

\bibitem{Manton:1981es}
  N.~S.~Manton,
  Nucl.\ Phys.\  B {\bf 193}, 502 (1981);
%
%
  G.~Chapline and R.~Slansky,
  Nucl.\ Phys.\  B {\bf 209}, 461 (1982);
%
  S.~Randjbar-Daemi, A.~Salam and J.~A.~Strathdee,
  Nucl.\ Phys.\  B {\bf 214}, 491 (1983);
%
  C.~Wetterich,
  Nucl.\ Phys.\  B {\bf 222}, 20 (1983);
%
  P.~H.~Frampton and K.~Yamamoto,
  Phys.\ Rev.\ Lett.\  {\bf 52}, 2016 (1984);
%
  P.~H.~Frampton and T.~W.~Kephart,
  Phys.\ Rev.\ Lett.\  {\bf 53}, 867 (1984);
%
  K.~Pilch and A.~N.~Schellekens,
  Nucl.\ Phys.\  B {\bf 256}, 109 (1985);

\bibitem{Witten:1984dg}
  E.~Witten,
  Phys.\ Lett.\  B {\bf 149}, 351 (1984).


\bibitem{Bachas:1995ik}
  C.~Bachas,
  arXiv:hep-th/9503030.

\bibitem{BDL}
  M.~Berkooz, M.~R.~Douglas and R.~G.~Leigh,
  Nucl.\ Phys.\  B {\bf 480}, 265 (1996)
  [arXiv:hep-th/9606139].

\bibitem{Blumenhagen:2000wh}
  R.~Blumenhagen, L.~Goerlich, B.~Kors and D.~Lust,
  JHEP {\bf 0010}, 006 (2000)
  [arXiv:hep-th/0007024].

\bibitem{Angelantonj:2000hi}
  C.~Angelantonj, I.~Antoniadis, E.~Dudas and A.~Sagnotti,
  Phys.\ Lett.\  B {\bf 489}, 223 (2000)
  [arXiv:hep-th/0007090].

\bibitem{CIM}
  D.~Cremades, L.~E.~Ibanez and F.~Marchesano,
  JHEP {\bf 0405}, 079 (2004)
  [arXiv:hep-th/0404229].

\bibitem{Troost:1999xn}
  J.~Troost,
  Nucl.\ Phys.\  B {\bf 568}, 180 (2000)
  [arXiv:hep-th/9909187].






\bibitem{Alfaro:2006is}
  J.~Alfaro, A.~Broncano, M.~B.~Gavela, S.~Rigolin and M.~Salvatori,
  JHEP {\bf 0701} (2007) 005
  [arXiv:hep-ph/0606070].



\bibitem{Aldazabal:2000dg}
  G.~Aldazabal, S.~Franco, L.~E.~Ibanez, R.~Rabadan and A.~M.~Uranga,
  J.\ Math.\ Phys.\  {\bf 42}, 3103 (2001)
  [arXiv:hep-th/0011073];
%
  JHEP {\bf 0102}, 047 (2001)
  [arXiv:hep-ph/0011132].

\bibitem{Blumenhagen:2000ea}
  R.~Blumenhagen, B.~Kors and D.~Lust,
  JHEP {\bf 0102}, 030 (2001)
  [arXiv:hep-th/0012156].

\bibitem{Cvetic:2001tj}
  M.~Cvetic, G.~Shiu and A.~M.~Uranga,
  Phys.\ Rev.\ Lett.\  {\bf 87}, 201801 (2001)
  [arXiv:hep-th/0107143];
%
  Nucl.\ Phys.\  B {\bf 615}, 3 (2001)
  [arXiv:hep-th/0107166];


\bibitem{Honecker:2004kb}
  G.~Honecker and T.~Ott,
  Phys.\ Rev.\  D {\bf 70}, 126010 (2004)
  [Erratum-ibid.\  D {\bf 71}, 069902 (2005)]
  [arXiv:hep-th/0404055];
%
%
  F.~Gmeiner and G.~Honecker,
  JHEP {\bf 0709}, 128 (2007)
  [arXiv:0708.2285 [hep-th]].



\bibitem{Blumenhagen:2005mu}
  R.~Blumenhagen, M.~Cvetic, P.~Langacker and G.~Shiu,
  Ann.\ Rev.\ Nucl.\ Part.\ Sci.\  {\bf 55}, 71 (2005)
  [arXiv:hep-th/0502005];
%
  R.~Blumenhagen, B.~Kors, D.~Lust and S.~Stieberger,
  Phys.\ Rept.\  {\bf 445}, 1 (2007)
  [arXiv:hep-th/0610327].


\bibitem{DiVecchia:2008tm}
  P.~Di Vecchia, A.~Liccardo, R.~Marotta and F.~Pezzella,
  arXiv:0810.5509 [hep-th].

\bibitem{Antoniadis:2009bg}
  I.~Antoniadis, A.~Kumar and B.~Panda,
  arXiv:0904.0910 [hep-th].

\bibitem{Abe:2009dr}
  H.~Abe, K.~S.~Choi, T.~Kobayashi and H.~Ohki,
  JHEP {\bf 0906}, 080 (2009)
  [arXiv:0903.3800 [hep-th]];
  Nucl.\ Phys.\  B {\bf 820}, 317 (2009)
  [arXiv:0904.2631 [hep-ph]].



\bibitem{Abe:2008fi}
  H.~Abe, T.~Kobayashi and H.~Ohki,
  JHEP {\bf 0809}, 043 (2008)
  [arXiv:0806.4748 [hep-th]];
  H.~Abe, K.~S.~Choi, T.~Kobayashi and H.~Ohki,
  Nucl.\ Phys.\  B {\bf 814}, 265 (2009)
  [arXiv:0812.3534 [hep-th]].



\bibitem{Abe:2009uz}
  H.~Abe, K.~S.~Choi, T.~Kobayashi and H.~Ohki,
  arXiv:0907.5274 [hep-th].


\bibitem{Kobayashi:2004ud}
T.~Kobayashi, S.~Raby and R.~J.~Zhang,
Phys.\ Lett.\ B {\bf 593}, 262 (2004)
[arXiv:hep-ph/0403065];
%
%
Nucl.\ Phys.\ B {\bf 704}, 3 (2005)
  [arXiv:hep-ph/0409098].

\bibitem{Forste:2004ie}
S.~Forste, H.~P.~Nilles, P.~K.~S.~Vaudrevange and A.~Wingerter,
Phys.\ Rev.\ D {\bf 70}, 106008 (2004);
%


\bibitem{Buchmuller:2005jr}
  W.~Buchmuller, K.~Hamaguchi, O.~Lebedev and M.~Ratz,
  Phys.\ Rev.\ Lett.\  {\bf 96}, 121602 (2006);
  Nucl.\ Phys.\  B {\bf 785}, 149 (2007)
  [arXiv:hep-th/0606187].


\bibitem{Kim:2006hw}
  J.~E.~Kim and B.~Kyae,
  Nucl.\ Phys.\  B {\bf 770}, 47 (2007)
  [arXiv:hep-th/0608086].
%

\bibitem{Lebedev:2006kn}
  O.~Lebedev, H.~P.~Nilles, S.~Raby, S.~Ramos-Sanchez, M.~Ratz, P.~K.~S.~Vaudrevange and A.~Wingerter,
  Phys.\ Lett.\  B {\bf 645}, 88 (2007)
  [arXiv:hep-th/0611095];
%
  Phys.\ Rev.\  D {\bf 77}, 046013 (2008)
  [arXiv:0708.2691 [hep-th]].



\bibitem{Blumenhagen:2005ga}
  R.~Blumenhagen, G.~Honecker and T.~Weigand,
  JHEP {\bf 0506}, 020 (2005)
  [arXiv:hep-th/0504232].



\bibitem{Choi:2009pv}
  K.~S.~Choi, T.~Kobayashi, R.~Maruyama, M.~Murata, Y.~Nakai, H.~Ohki and M.~Sakai,
  arXiv:0908.0395 [hep-ph].

\bibitem{toron}
  G.~'t Hooft,
  Nucl.\ Phys.\  B {\bf 153}, 141 (1979); 
  Commun.\ Math.\ Phys.\  {\bf 81} (1981) 267;
  P.~van Baal,
  Commun.\ Math.\ Phys.\  {\bf 94} (1984) 397;
  Z.~Guralnik and S.~Ramgoolam,
  Nucl.\ Phys.\  B {\bf 521}, 129 (1998)
  [arXiv:hep-th/9708089].



\bibitem{Antoniadis:2004pp}
  I.~Antoniadis and T.~Maillard,
  Nucl.\ Phys.\  B {\bf 716}, 3 (2005)
  [arXiv:hep-th/0412008].


\bibitem{Bourjaily:2009ci}
  J.~L.~Bourjaily,
  arXiv:0905.0142 [hep-th].


\bibitem{Salvatori:2006pb}
  M.~Salvatori,
  JHEP {\bf 0706}, 014 (2007)
  [arXiv:hep-ph/0611309];
%
%
  D.~Hernandez, S.~Rigolin and M.~Salvatori,
  arXiv:0712.1980 [hep-ph];
%
  G.~von Gersdorff,
  Nucl.\ Phys.\  B {\bf 793}, 192 (2008)
  [arXiv:0705.2410 [hep-th]];
%
%
%
  A.~F.~Faedo, D.~Hernandez, S.~Rigolin and M.~Salvatori,
  arXiv:0911.0997 [hep-ph].


\bibitem{Abe:2010ii}
  H.~Abe, K.~S.~Choi, T.~Kobayashi and H.~Ohki,
  arXiv:1001.1788 [hep-th].



\end{thebibliography}
\end{document}